%% file: SEAMS.tex
\documentclass[10pt,conference]{IEEEtran}

\IEEEoverridecommandlockouts
\usepackage[T1]{fontenc}
\usepackage{amsmath,amssymb,amsfonts}
\usepackage{algpseudocode}
\usepackage{algorithm}
\usepackage{graphicx}
\usepackage{textcomp}
\usepackage{xcolor}
\usepackage{url}
\usepackage{subcaption}
\usepackage{listings,xcolor}
\usepackage{adjustbox}
\usepackage{booktabs}  % for better lines
\usepackage{array}     % for column formatting
\usepackage{longtable} % in case you have multi-page tables
\usepackage{soul}

\usepackage[numbered]{matlab-prettifier}%break listing by default

\colorlet{punct}{red!60!black}
\definecolor{background}{HTML}{EEEEEE}
\definecolor{delim}{RGB}{20,105,176}
\colorlet{numb}{magenta!60!black}

\lstdefinelanguage{JSON}{
    basicstyle=\tiny\ttfamily,
    numbers=left,
    numberstyle=\scriptsize,
    stepnumber=1,
    numbersep=8pt,
    showstringspaces=false,
    string=[s]{"}{"},
    breaklines=true,
    frame=lines,
    backgroundcolor=\color{background},
    string=[s]{"}{\"},
    literate=
     *{0}{{{\color{numb}0}}}{1}
      {1}{{{\color{numb}1}}}{1}
      {2}{{{\color{numb}2}}}{1}
      {3}{{{\color{numb}3}}}{1}
      {4}{{{\color{numb}4}}}{1}
      {5}{{{\color{numb}5}}}{1}
      {6}{{{\color{numb}6}}}{1}
      {7}{{{\color{numb}7}}}{1}
      {8}{{{\color{numb}8}}}{1}
      {9}{{{\color{numb}9}}}{1}
      {:}{{{\color{punct}{:}}}}{1}
      {,}{{{\color{punct}{,}}}}{1}
      {\{}{{{\color{delim}{\{}}}}{1}
      {\}}{{{\color{delim}{\}}}}}{1}
      {[}{{{\color{delim}{[}}}}{1}
      {]}{{{\color{delim}{]}}}}{1},
}

\def\BibTeX{{\rm B\kern-.05em{\sc i\kern-.025em b}\kern-.08em
    T\kern-.1667em\lower.7ex\hbox{E}\kern-.125emX}}

\newcommand{\changed}[1]{\textcolor{black}{#1}}
\newtheorem{definition}{Definition}

\begin{document}

\title{
Adaptive Human-Robot Collaborative Missions using Hybrid Task Planning
% Adaptive
% Adaptation with Retries for Human-Robot Coordination
% \thanks{Funding agency: AI4Work, Europe Horizon*.}
}

% ANONYMUS for submission
\author{
\IEEEauthorblockN{Gricel V\'{a}zquez, Alexandros Evangelidis, Sepeedeh Shahbeigi, Simos Gerasimou}
\IEEEauthorblockA{\textit{University of York, UK}}
% \textit{University of York, UK}\\
\IEEEauthorblockA{\{gricel.vazquez,alexandros.evangelidis,sepeedeh.shahbeigi,simos.gerasimou\}@york.ac.uk}
% \and
% \and
% \and
% \and
% \and
}

\maketitle

\begin{abstract}
% \textcolor{red}{Change motivation that hybrid is better - Heuristic planning solvers do not consider uncertainty and PMC exploits. }

Producing robust task plans in human-robot collaborative missions is a critical activity in order to increase the likelihood of these missions completing successfully. 
Despite the broad research body in the area, which considers different classes of constraints and uncertainties, its applicability is confined to relatively simple problems that can be comfortably addressed by the underpinning mathematically-based or heuristic-driven solver engines. 
In this paper, we introduce a hybrid approach that effectively solves the task planning problem by decomposing it into two intertwined parts, starting with the identification of a feasible plan and followed by its uncertainty augmentation and verification yielding a set of Pareto optimal plans. To enhance its robustness, adaptation tactics are devised for the evolving system requirements and agents' capabilities. We demonstrate our approach through an industrial case study involving workers and robots undertaking activities within a vineyard, showcasing the benefits of our hybrid approach both in the generation of feasible solutions and scalability compared to native planners.

% \textcolor{red}{\url{https://conf.researchr.org/track/seams-2025/seams-2025-research-track#Submission}}

\end{abstract}

\begin{IEEEkeywords}
hybrid task planning, multi-robot multi-human systems, genetic algorithms
\end{IEEEkeywords}

\section{Introduction \label{sec:introduction}}

Cyber-Physical-Human Systems (CPHS) are advanced, interconnected systems comprising human agents, robotic components, and cyber infrastructure~\cite{annaswamy2023cyber}. These systems are increasingly used in diverse domains such as agriculture~\cite{tokekar2016sensor}, manufacturing~\cite{lee2015manufacturing}, and healthcare~\cite{calinescu2019socio}, where they must collaborate effectively to complete their tasks while adhering to functional and non-functional, probabilistic and non-probabilistic requirements.  A crucial CPHS challenge is developing robust task plans that account for uncertainty and ensure the successful completion of missions under dynamically changing conditions, while ensuring \changed{plan} correctness~\cite{sanchez2024automated,camara2024uncertainty,weyns2023towards,zhao2024bayesian}. 

Uncertainty in CPHS arises from various factors, including unpredictable human behaviour, environmental variability, and mechanical failures~\cite{man2016uncertainty}. Addressing such uncertainty often requires the verification of task plans by considering probabilistic properties, such as the likelihood of mission success. Probabilistic Model Checking (PMC) is a widely recognized approach for verifying these properties~\cite{calinescu2012adaptive,kwiatkowska2010advances}. However, while PMC excels at providing formal guarantees, its computational requirements grow exponentially with the size and complexity of the problem, a limitation commonly referred to as the state explosion problem~\cite{kwiatkowska2007stochastic}. This makes PMC impractical for large-scale CPHS task planning scenarios involving numerous agents, tasks, and potential disruptions~\cite{kwiatkowska2022probabilistic,gerasimou2021evolutionary}.

Existing approaches to task planning in CPHS can be broadly classified into deterministic and probabilistic methods~\cite{fischer2021loop}. Deterministic planners, such as those based on classical planning models like PDDL~\cite{younes2004ppddl1}, are computationally efficient but fail to address uncertainties arising from dynamic environments or probabilistic task outcomes~\cite{calinescu2017using}. Probabilistic planning frameworks, such as those using Markov Decision Processes (MDPs), provide a formal way to model uncertainty but suffer from scalability issues due to the exponential growth of state space, as seen in large-scale multi-agent systems~\cite{kwiatkowska2022probabilistic}. Probabilistic model checking tools like PRISM~\cite{kwiatkowska2011prism} and Storm~\cite{hensel2022storm} have been extensively used for verification, yet their application is often limited to relatively small problem instances due to computational constraints.

Recent advanced hybrid approaches combine the strengths of numerical and probabilistic methods. For instance, approaches like~\cite{camara2020haiq,vazquez2022scheduling} integrate constraint solving with probabilistic verification, while probabilistic extensions of the main definition language (PPDDL)~\cite{younes2004ppddl1} facilitate modelling and verification of uncertain planning domains. Despite these efforts, there remains a need for solutions that can generate adaptable, verified plans capable of addressing real-world requirements, such as task retries, energy consumption, and human factors, while accommodating runtime changes.

In this paper, we propose a hybrid task planning approach that decomposes the planning problem into two stages to tackle the scalability and efficiency challenges. First, we employ off-the-shelf numerical planning techniques to generate a feasible initial plan. Next, we augment this plan with uncertainty \changed{information yielding a parametric probabilistic model. Meta-heuristic search allows us to synthesise Pareto-optimal task plans that meet probabilistic requirements.} We also introduce a task planning adaptation algorithm for the generation and unfolding of new correct and verified plans, if required.  

We evaluate our approach in an industrial case study involving task execution at a vineyard, where human workers and robots collaborate to perform vineyard activities. The results demonstrate that our hybrid framework effectively balances the scalability of numerical planning with the rigour of probabilistic verification. By supporting incremental adaptation to runtime changes, such as task failures or evolving requirements, our approach ensures both robustness and flexibility in the CPHS plan, while \changed{also reducing computational demands}.

Our main contributions are as follows: 
(1)~we present a hybrid approach for the generation of correct verified plans leveraging numerical planning and PMC to mitigate the state explosion problem; 
(2)~we integrate meta-heuristic search to synthesise  Pareto-optimal plans concerning maximising the probability of mission success while minimising the mission cost; 
(3) we propose an algorithm for task plan adaptation in response to failures in the completion of tasks and changes in requirements; and 
(4)~we evaluate our hybrid planning approach and the adaptation algorithm on an industrial vineyard case study provided by our project partners.

Paper structure. 
Section~\ref{sec:example} describes our industrial case study. Section~\ref{sec:background} provides the required background. Section~\ref{sec:ProblemFormulation} describes the problem formulation,  while Sections~\ref{sec:approach} and~\ref{sec:evaluation} present our approach and its evaluation, respectively. Finally, Sections~\ref{sec:relatedWork} and \ref{sec:conclusions} cover related work and conclusions.

\begin{figure}[]%tbp]
\centering
\begin{minipage}[t]{1\linewidth}
\begin{subfigure}[b]{0.55\textwidth} 
  \centering
  \includegraphics[width=\linewidth]{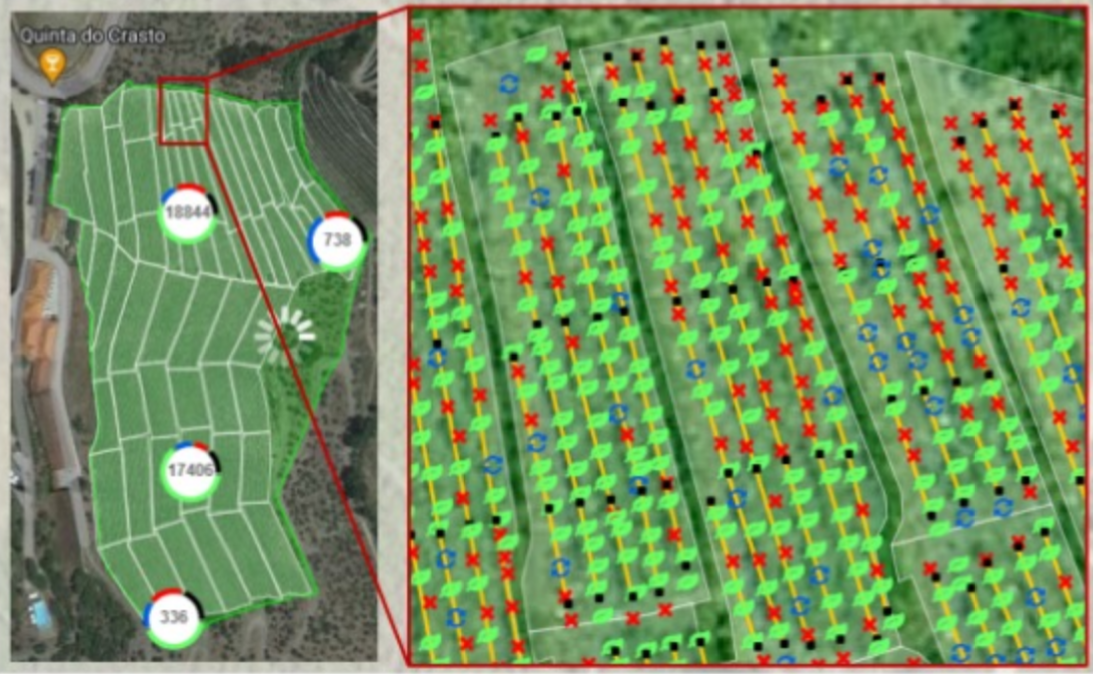}
  \caption{Vineyard layout.}
  \label{fig:vineyard}
\end{subfigure}%
\hfill%
\begin{subfigure}[b]{0.4\textwidth} 
  \centering
  \includegraphics[width=0.8\linewidth]{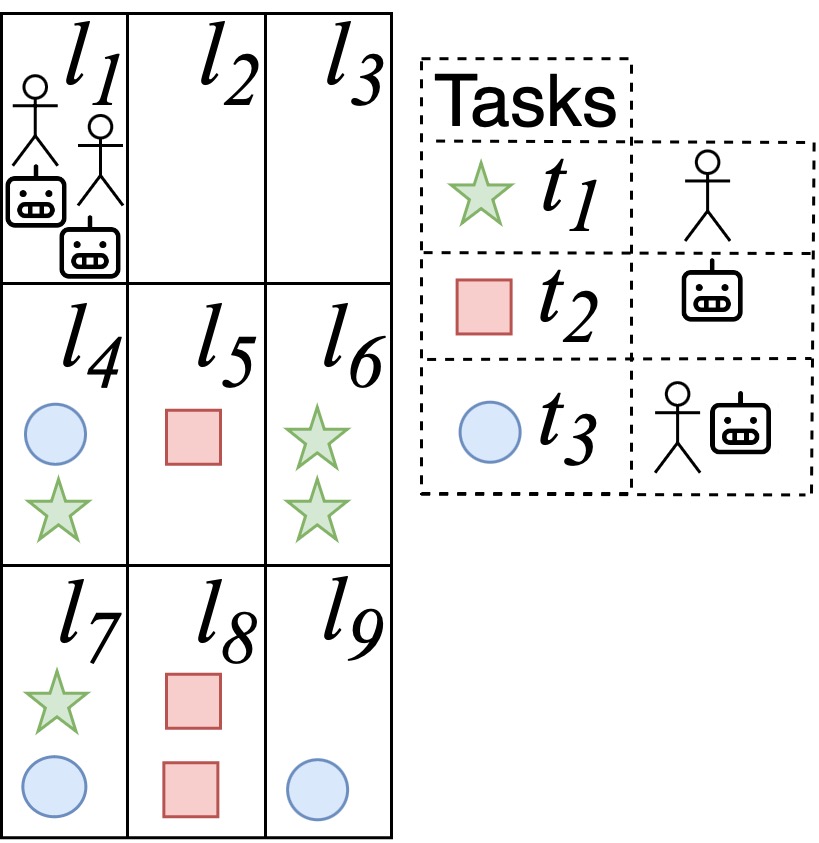}
  \caption{World abstraction}
  \label{fig:worldAbst}
\end{subfigure}
\caption{Vineyard layout and abstract representation. Locations $l_1$-$l_9$ represent a section of a vineyard's row, where tasks $t1$-$t3$ have to be completed. Task $t1$ can be done by human workers, $t2$ by robots and $t3$ by either.}
\label{fig:test}
\end{minipage} 
\vspace{-4mm}
\end{figure}

%\vspace{-1mm}
\section{Vineyard Inspection CPHS Mission
\label{sec:example}
}

We motivate our approach through an industrial case study \changed{on task automation in a vineyard in Portugal, provided by our research partner Quinta Do Castro and the University of Trás-os-Montes} (see Figure~\ref{fig:vineyard}). The vineyard is divided into subsections, each with multiple rows where different varieties of vines grow. 
The old vineyards are equipped with GPS technology, providing the precise location of each plant. 

The vineyard tasks are classified as follows:

    \noindent $\bullet$
    Harvesting ($t1$): Grape harvesting is an extremely arduous and repetitive activity. As a highly-dexterous task, this is delegated only to \changed{humans}. Severe fatigue has to be mitigated whenever possible as, after \changed{many} days in steep terrain conditions, workers show signs of physical and mental fatigue~\cite{wang2022toward}. 
    
    % \item 
    \noindent $\bullet$
    Vineyard monitoring ($t2$): Continuous monitoring of vineyards includes monitoring the vines' health, the vines' state after a heatwave or heavy rainfall, and routine monitoring. As remote areas can be extremely difficult for humans to reach and magnify human fatigue, this task is delegated to robots. 

    % \item 
    \noindent $\bullet$
    Grapevine identification ($t3$): Grapevine varietal identification occurs during the observation period between the hot months of June and August. This task, comprising vine leaf pictures taken and sent to a central unit, is shared between human workers and robots. This task might fail due to communication errors, hardware failures, or the collection of unsatisfactory low-quality images.
% \end{itemize}

Costs and success probabilities associated with each task are shown in Table~\ref{table:vineyard_costs_prob}. Costs are representative of human fatigue and the robot's battery consumption, both scaled from 1 to 5 units. 
We assume that tasks can be retried up to 5 times by human workers and up to 10 times by robots. 
Tasks are located at different locations ($l_1$-$l_9$) as shown in Figure~\ref{fig:worldAbst}. Human workers ($w1, w2$) and robots ($r1, r2$) start at $l_1$ and can only move between adjacent locations. They must coordinate to complete all tasks while avoiding each other at all times after deployment. Each move incurs a cost of 1 unit and is allowed only if the destination location is unoccupied.

%The whole mission must be completed within four hours. 

\begin{table}[t]
\caption{Task cost, success probability and maximum number of retries per task and agent (human worker or robot).}
\vspace{-2mm}
\centering
\begin{tabular}{cccc|ccc|c}
\cline{2-8}
 & \multicolumn{3}{c|}{Cost} & \multicolumn{3}{c|}{Success prob.} & Max. retries \\ \hline
\multicolumn{1}{c|}{Task} & $t1$ & $t2$ & $t3$ & $t1$ & $t2$ & $t3$ & $t1,t2,t3$ \\ \hline
\multicolumn{1}{c|}{Human} & 3 & - & 5 & 1.00 & - & 0.99 & 5 \\
\multicolumn{1}{c|}{Robot} & - & 1 & 1 & - & 0.99 & 0.97 & 10 \\ \hline
\end{tabular}
\label{table:vineyard_costs_prob}
\vspace{-4mm}
\end{table}

Addressing this problem entails producing a plan that enables successful mission completion while minimising human fatigue and energy costs. 
During operation, tasks might fail or requirements might change and the system is required to recover whenever possible to mitigate disruptions. Hence, maximising the probability of success while minimising cost and time is required. Finally, the task planner must ensure a success probability of 0.95.

\section{Background
\label{sec:background}
}

\subsection{Numeric planning}\label{sec:numericPlanning}

A numerical planning model consists of \changed{a} domain, specifying the world and its behaviours, and \changed{a} planning problem, describing the specific task to be solved within that world~\cite{ghallab2016automated, fox2003pddl2}. 
%~(inspired by ~\cite{lindsay2023using}):
The tuple $\mathbb{D}=(\mathbb{T},\mathbb{P},\mathbb{F},\mathbb{A})$ defines a domain, where $\mathbb{T}$ is a set of types, $\mathbb{P}$ a set of predicates, $\mathbb{F}$ a set of numerical-valued functions, and $\mathbb{A}$ a set of actions. Types categorise objects in the domain; predicates represent properties or relationships between objects; and actions describe possible state transitions, each comprising preconditions and effects. 
Functions represent quantities (e.g., distance, fuel level,  travelling cost) and might depend on a finite number of typed objects. %(i.e., objects created from types). 
A numerical fluent is a function that models quantities that change over time.

A planning problem is defined by the tuple $\mathbf{P}=(\mathbf{O},s_{0},g,o)$ where $\mathbf{O}$ is the set of typed objects, $s_{0}$ is a set of grounded predicates that define the state of the world at the initial state, $g$ is a set of predicates that must be satisfied in the goal state, and $o$ is the planning optimisation objective~\cite{fox2003pddl2}. 
Objects are instances of domain types.

\textbf{PDDL}. The planning model is aligned with the standard Planning Domain Description Language (PDDL)~\cite{younes2004ppddl1}. 
Different variants exist to accommodate classical, numeric or temporal planning requirements. 
The PDDL requirements for our task planning problem without uncertainty quantification are \textit{:requirements :strips :typing :negative-preconditions :numeric-fluents}. 
Hence, we use PDDL2.1~\cite{fox2003pddl2}, which allows conditional statements (e.g. $x>0.5$) and numerical fluents (e.g., \(x+=1\)) in preconditions and effects, respectively. 
We refer interested readers to~\cite{fox2003pddl2} for a full description.%the complete syntax.% (and further description of the domain and planning problem parts).

\textbf{Heuristic numerical planning}. 
\changed{The Expressive Numeric Heuristic Search Planner (ENHSP)~\cite{scala2016interval} is a planner that can solve problems defined in PDDL.}
ENHSP uses an expressive representation for the planning problem and the heuristic search process and integrates numerical fluents to handle complex numeric constraints and effects efficiently. 
By combining domain-specific heuristics and search techniques, ENHSP can find optimal or near-optimal solutions more effectively than traditional planners in domains with numerical variables.

\subsection{Probabilistic Model Checking}
Probabilistic model checking (PMC)~\cite{kwiatkowska2007stochastic} is a formal verification technique for the quantitative analysis of probabilistic systems.
PMC tools, such as PRISM \cite{kwiatkowska2011prism} and Storm \cite{hensel2022storm} automatically verify properties of such systems.

% \subsubsection{Probabilistic models (DTMCs/MDPs)}
\vspace{1mm}\noindent
\textbf{Discrete-time Markov chain (DTMC).}
A DTMC is a tuple $\mathcal{D} = (S, \bar{s}, \delta, AP, L)$ where $S$ is a finite set of states and $\bar{s} \in S$ is an initial state; $\delta : S \rightarrow \mathit{Dist}(S)$ is a probabilistic transition function, mapping states to probability distributions over $S$; $AP$ is a set of atomic propositions; and $L : S \rightarrow 2^{AP}$ is a state labelling function.

\textbf{Markov Decision Process (MDP).}  
An MDP extends a DTMC with actions, formalized as $\mathcal{Q} = (S, \bar{s}, Act, \delta, AP, L)$, where $Act$ is a finite action set and $\delta : S \times Act \rightarrow \mathit{Dist}(S)$ is a partial transition function. For $s \in S$, let $Act(s) = \{a \in Act \mid \delta(s,a) \text{ is defined}\}$, with $Act(s) \neq \emptyset$.

\vspace{1mm}\noindent
\textbf{Property Specification.}
To specify and analyse properties of probabilistic systems, we use quantitative extensions of temporal logic.
%In particular, we use  the PRISM property specification language, which extends  Probabilistic Computation Tree Logic (PCTL) augmented with reward-based operators~\cite{kwiatkowska2022probabilistic,kwiatkowska2011prism}.
In particular, we use  Probabilistic Computation Tree Logic (PCTL) augmented with reward-based operators and leverage the PRISM property specification language~\cite{kwiatkowska2022probabilistic,kwiatkowska2011prism}.
Given the set $AP$ of atomic propositions, we identify four types of constructs: 
(i) \emph{state formulae:} 
$\phi ::= \textit{true} \mid\ a \mid\; \neg \phi \;\mid\; \phi \land \phi,$ 
where $a \in AP$; 
(ii) \emph{ path formulae:} 
$\psi ::= \mathsf{X}\,\phi \;\mid\; \phi\,\mathsf{U}^{\le k}\,\phi \;\mid\; \phi\,\mathsf{U}\,\phi$; 
(iii)\emph{ reward formulae:} 
$\rho ::= \mathcal{I}^{=k} \mid \mathcal{C}^{\le k} \mid \mathcal{F}\,\phi$;
and (iv) \emph{queries:} %\emph{properties:} 
$\Phi ::= \mathcal{P}_{\bowtie p} [\psi] \;\mid\; \mathcal{R}^r_{\bowtie q} [\rho],$ 
where $\bowtie \in \{\!<,\le,>,\ge\!\}$, $p \in [0,1]$, $r$ is a reward structure, $q \in \mathbb{R}_{\ge 0}$, and $k \in \mathbb{N}$.

For example, the \emph{quantitative} query $\mathcal{P}_{=?}[\mathcal{F} \text{"success"}]$ computes the probability of eventually reaching a state labelled "success", while  \(\mathcal{R}_{=?}[\mathcal{F} \text{"success"}]\) computes the corresponding expected reward. 
\emph{Qualitative} queries check threshold conditions; for example, 
$\mathcal{P}_{\ge 0.9}[\mathcal{F} \text{"success"}]$ checks if the probability to reach a "success" state is at least 0.9. 
For MDPs, \emph{min/max queries}, like $\mathcal{P}_{min/max = ?}$, and their reward-based counterparts $\mathcal{R}_{min/max =?}$ enable computing the minimum/maximum probability (or reward) over all MDP policies.

%\subsubsection{Probabilistic Temporal Logic with Rewards (rPCTL)}
%A probabilistic model (DTMC or MDP) can be augmented with a reward structure $r_s = (r_S, %r_A)$ where $r_S: S \rightarrow \mathbb{Q}$ is a state reward function mapping states to %rational values and  $r_A: S \times A \rightarrow \mathbb{Q}$ is an action reward function assigning values to state-action pairs (undefined for DTMCs). The reward-based properties extend the PCTL syntax with formulae of the form $\mathcal{R}_{\bowtie r}[\rho]$, where ${\bowtie} \in \{<, \leq, >, \geq\}$, $r \in \mathbb{R}_{\geq 0}$, and $\rho$ is one of the following reward operators: $\mathcal{I}^{=k}$, denoting instantaneous reward at time step $k$, $\mathcal{C}^{\leq k}$ and $\mathcal{F} \phi$, representing cumulative rewards up to time step $k$ and reachability rewards until satisfying $\phi$, respectively. \gricel{missing PCTL syntax. Could we change it to PRISM language? As this allows also for quantitative queries: $\mathcal{R}$=?[$\mathcal{F}$ done], P=?[F done]}

\begin{figure*}[t]
\centering
  \includegraphics[width=0.95\textwidth]{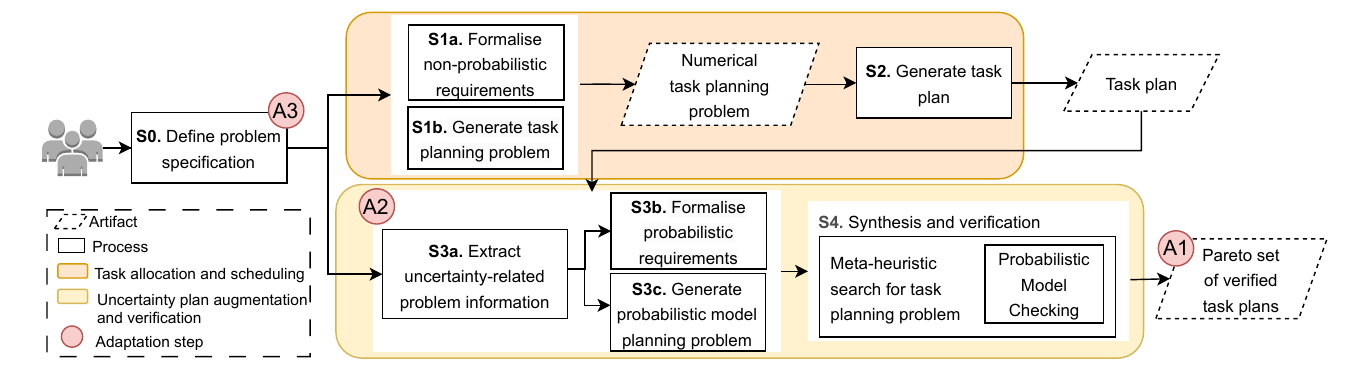}
  \vspace{-2mm}
  \caption{Overview of our approach showing the required steps and generated artefacts, and adaptation relevant parts.}
  \label{fig:overview}
  \vspace{-4mm}
\end{figure*} 

\section{Problem Formulation}
\label{sec:ProblemFormulation}
We consider a CPHS \changed{comprising a} set of agents (robots and humans) $\mathcal{A}=\{a_1, a_2, a_3, \ldots\}$ that reside in a world with locations $\mathcal{L}=\{l_1, l_2, l_3, \ldots\}$. \changed{A} predicate $\mathit{Path}: {\mathcal{L} \times \mathcal{L} \to \mathbb{B}}$ indicates whether a path exists between location pairs with distance $dist_{l_i,l_j} \in \mathbb{R}$. 
The mission entails \changed{a set of} tasks $\mathcal{T}=\{t_1, t_2, \ldots, \}$, \changed{each to be completed at a given location $\mathit{TaskLoc}: \mathcal{T} \times \mathcal{L} \to \mathbb{B}$.}
Each agent can perform actions $\mathcal{A}ct = \{Move, Do\}$, where $Move: \mathcal{A}\times \mathcal{L} \times \mathcal{L} \to \mathbb{B}$ and $Do: \mathcal{A} \times \mathcal{T} \times \mathcal{L} \to \mathbb{B}$.
The probability of an agent successfully performing a task is given by  $PSuccess:\mathcal{A}\times \mathcal{T} \to [0,1]$, where $0$ means that the agent cannot perform the task.
In case of a failure, an agent can retry the task up to a maximum number of possible retries given by $Retry:\mathcal{A}\times \mathcal{T} \to \mathbb{N}_0$. 
Finally, the cost incurred by an agent to perform a task is given by $CostT:\mathcal{A}\times \mathcal{T} \to \mathbb{R}_0^+$.
These functions neatly capture differences between the agents; for example, $PSuccess$ allows specifying that humans are more competent in dexterous tasks, such as picking grapes, requiring fewer retries than a robot, while $CostT$ allows specifying the robots' energy consumption or the increase in a human's fatigue. The CPHS mission includes the constraints $C$ 
% $C= \{C_1, \ldots, C_5\}$ 
and optimisation objectives~$O$
% $O= \{O_1, O_2\}$ 
shown in Table~\ref{table:missionReqs}\changed{. A}n optimisation objective defines if a quality attribute should be maximised or minimised provided that all constraints are met.

\begin{table}[t]
\caption{Mission requirements partitioned into constraints $C_i$ and optimisation objectives $O_j$.}
\vspace{-2mm}
\label{table:missionReqs}
\begin{tabular}{p{0.1cm} %p{1cm} 
                p{7.7cm} }
\toprule
    \textbf{ID} 
        % & \textbf{Type} 
        & \textbf{Description} \\ \hline
    \textbf{$C_1$}
        % & Constraint 
        &An agent can travel between locations iff a path exists.\\
    \textbf{$C_2$}
        & An agent must be at the initial location of the path \changed{to travel}. \\
    \textbf{$C_3$}
        & Each location must be occupied by at most one agent at any time. \\
    \textbf{$C_4$}
        & Agent $a$ can be allocated task $t$ iff $PSuccess(a,t)\! \geq\!\gamma, \gamma\!\in\!(0,1]$\\
    \textbf{$C_5$}
        & A plan is feasible iff all tasks $\mathcal{T}$ are completed.\\ 
        % and the overall mission success probability is at least $p \in (0,1]$\\
    \textbf{$C_6$}
        & A plan is feasible iff the probability of succeeding is at least $p_{succ}$.\\ 
    \hline
    \textbf{$O_1$}
        & The mission execution cost must be minimised.\\
    \textbf{$O_2$}
        & The overall mission success probability must be maximised.\\
\bottomrule
\end{tabular}
\vspace{-4mm}
\end{table}

\vspace{1mm}
\begin{definition}[Problem Specification] \label{def:probSpec}
A CPHS planning problem is defined by the tuple 
$\mathcal{M} = (\mathcal{A}, \mathcal{L}, \mathcal{T},\mathcal{A}ct, C, O, s_0$), where 
$\mathcal{A}$ is a set of agents,
$\mathcal{L}$ is a set of locations,
$\mathcal{T}$ is a set of tasks, 
$\mathcal{A}ct$ is a set of actions, 
$C$ and $O$ are the sets of mission constraints and optimisation objectives, respectively, and $s_0$ is the initial CPHS configuration comprising the set of pending tasks and the initial location of the agents.
\end{definition}

% \vspace{1mm}
% \begin{definition}[Feasible Plan]
% % Given the problem specification $\mathcal{M}$, 
% A feasible solution to the problem $\mathcal{M}$ entails is a plan $\Pi = \prod_{a \in \mathcal{A}}\prod_{1 \ldots H} \mathcal{A}ct_a$ comprising the set of actions performed per agent $a$ during the mission's horizon $H$ such that the mission constraints $C$ are satisfied, i.e., $\bigwedge_{c \in C} B(\Pi, c)$, where $B(\Pi, c) \in \mathbb{B}$ is \emph{True} if constraint $c$ is satisfied by plan $\Pi$ and \emph{False}, otherwise.
% \end{definition}

\vspace{1mm}
\begin{definition}[Problem Solution]
A solution to the planning problem $\mathcal{M}$ is a plan comprising the \changed{list} of actions performed per agent $a$ during \changed{its} (finite) execution horizon \changed{$H_a$} of the mission, represented by 
%$\Pi = \prod_{a \in \mathcal{A}}\prod_{1 \ldots H_a} \mathcal{A}ct_a$.
\changed{$\Pi = \left( \left(\mathcal{A}ct_a^h\right)_{h \in 1 \ldots H_a}\right)_{a \in \mathcal{A}}$}.
\end{definition}

\vspace{1mm}
\begin{definition}[Pareto Front Solutions]
Given the set of plans $\Pi^S$, the planning problem comprises finding the Pareto-optimal front $PF$ induced by the Pareto-optimal set $PS$ of plans that satisfy the $C$ constraints and are Pareto-optimal with respect to the $O$ optimisation objectives. Formally:
\begin{equation}\label{eq:ps}
    PS = \{ \Pi \in \Pi^S | \bigwedge_{c \in C} B(\Pi, c) \land (\nexists\Pi' \in \Pi^S \bullet \Pi' \prec \Pi)  \}
\end{equation}
\begin{equation}\label{eq:pf}
    PF = \{(o,\Pi)_{o \in O} | \Pi \in PS\}
\end{equation}
where $B(\Pi, c) \in \mathbb{B}$ is \emph{True} if constraint $c$ is satisfied by plan $\Pi$ and \emph{False}, otherwise, and $\prec : \Pi^S \times \Pi^S \to \mathbb{B}$ is the conventional dominance relation from Pareto optimisation~\cite{coello2006evolutionary}.
\end{definition}

%%%%%%%%%%%%%%%%%%%%%%%%%%%%%%%
%%%%%%%%%%%%%%%%%%%%%%%%%%%%%%%
%%%%%%%%%%%%%%%%%%%%%%%%%%%%%%%
\section{Adaptive and Hybrid Task Planning}
\label{sec:approach}

% \noindent 
\subsection{Overview}
Our hybrid task planning approach realises a separation of concerns between the probabilistic and non-probabilistic problem elements of $\mathcal{M}$ to address its intrinsic complexity and scalability issues, and solve\changed{s} equations~\eqref{eq:ps}~and~\eqref{eq:pf}. 
Fig.~\ref{fig:overview} shows the key steps of our approach. 
Step \emph{S0} involves defining the problem specification $\mathcal{M}$, typically provided by engineers, with input from stakeholders and domain experts.
In step \emph{S1}, \changed{this} problem specification is transformed into a task planning problem and its non-probabilistic requirements are formalised. 
In step \emph{S2}, a feasible plan is generated using non-probabilistic numeric planners and constraint solvers.
Next, step \emph{S3} extracts uncertainty-relevant information from the problem specification to both devise probabilistic-related requirements and augment the plan produced in step \emph{S2} with such information (e.g., number of retries, the expected probability of success).
Finally, step \emph{S4} employs meta-heuristics to synthesise the Pareto front \changed{(}PF\changed{)} and the corresponding Pareto set (PS\changed{)} of revised plans that are robust to uncertainties in the CPHS mission.
% identify task retry combinations that maximise success probability while minimising cost. 
During this step, probabilistic models of the task planning problem are automatically generated and formally verified using PMC. 
At runtime, our approach enables incremental adaptation in response to CPHS changes by selecting an alternative plan from the Pareto set (A1), re-executes the probabilistic synthesis (A2) if the original plan is still feasible and only resorts to adaptation from scratch (A3) if it cannot deal with the changes or guarantee a new set of constraints.

\subsection{Approach}
%\noindent
\textbf{Problem Specification (\emph{S0}):} Starting at step \emph{S0}, our approach involves defining the problem specification $\mathcal{M}$, encoding information needed for the CPHS mission. 
This model $\mathcal{M}$ conforms to the JSON format.
The following example presents its structure for the vineyard CPHS mission from Section~\ref{sec:example}. 
% is formatted using the JSON notation. 
% We use the following example to explain its sywith syntax explained with the following example. 
% the next example based on our vineyard case study.
%It contains the set of locations, tasks and agents $L, T, A \in \mathcal{M'}$; existing paths $Path\in F$; task locations $Taskloc\in F$; task costs $CostT\in F$, possible number of retries $Retry\in F$ and probabilities of success $PSuccess\in F$ associated to agents. Constraints of minimum probability of mission success $\Phi_5$ and minimum probability of assignment $\Phi_4$, $\Phi_{1,2}\in\Phi$. 

\begin{figure*}[t]
\centering
  \includegraphics[width=\textwidth, height=0.446\textheight]{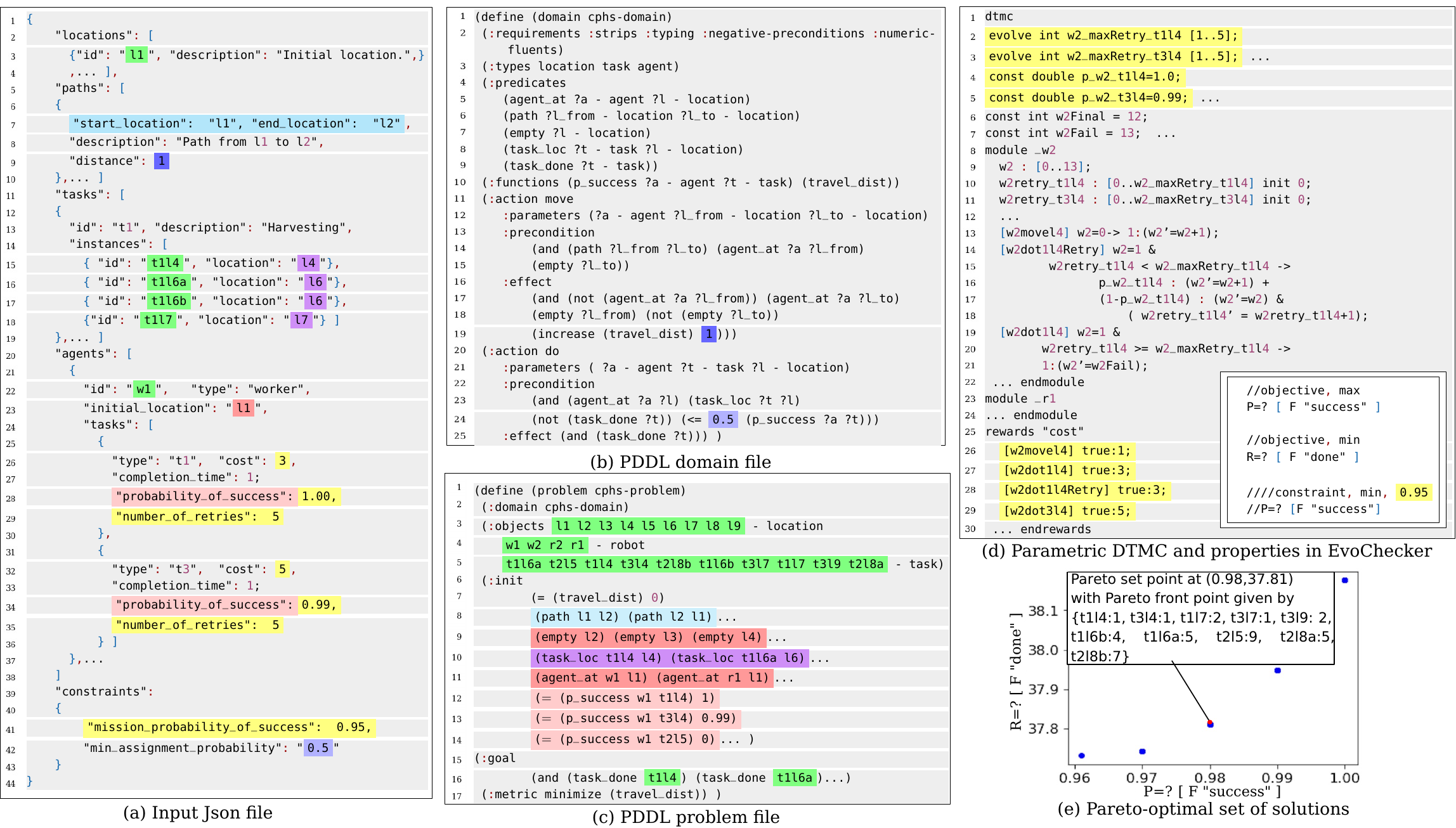}
  \vspace{-4mm}
  \caption{Input file~(a), generated PDDL files (b,c), parametric DTMC with probabilistic properties (d) and synthesis Pareto-optimal set of solutions (plans). The highlighted sections of the PDDL files use a colour palette based on the corresponding parts of the input file. The yellow parts refer to the task planning with uncertainty quantification part (see Figure~\ref{fig:overview}).}
  \label{fig:inputJsonFile}
  \vspace{-4mm}
\end{figure*}

\vspace{0.5mm}%\noindent  
\emph{Example 1 (Problem Specification).} Figure~\ref{fig:inputJsonFile}a shows the syntax of the problem specification file for our vineyard case study. 
This is divided into locations $\mathcal{L}$, paths, tasks $\mathcal{T}$, agents $\mathcal{A}$, and mission constraints $C$ and objectives $O$. 
Location elements comprise a unique identifier %\changed{($l\in \mathcal{L}$)} 
and an optional description. 
Path objects contain start and end location identifiers, an optional description, and a travelling cost ($distance=1$).
% \footnote{For simplicity, define a single travel cost for all costs.} 
Task objects define groups of inter-related tasks, comprising a unique identifier (e.g., $t1$), an optional description and a list of tasks with identifiers ($t1l4,\ t1l6a,\ t1l6b,\ t1l7\in \mathcal{T}$) and locations (e.g., $Taskloc(t1l4)=l4$). %The design decision to group tasks simplifies the definition of agent objects as follows. 
Agent objects are defined by a unique identifier, a type, $type\in\{\text{``worker"}, \text{``robot"}\}$, and a set of tasks the agent can perform. 
Each task has an identifier (e.g., $t1$ which refers to tasks $t1l4,t1l6a,t1l6b$ \changed{and} $t1l7$), an expected \changed{cost} (e.g., $\mathit{CostT}(t1l4)=3$), a probability of success (e.g., $\mathit{PSuccess}(t1l4)=1$) and the number of retries allowed per agent (e.g., $\mathit{Retry}(t1l4)=5$). 
In our example, $worker1$ can perform tasks $t1$ and $t3$. 
Lastly, two constraints describe the probability of success ($p_{\text{succ}}$) given by $mission\_probability\_of\_success=\changed{0.95}$, and the allocation threshold $\gamma = 0.5$ (``min\_assignment\_probability'').

%see https://sebastianhahner.de/publications/2024/Camara2024_UncertaintyFlowDiagramsTowardsASystematicRepresentationOfUncertaintyPropagationAndInteractionInAdaptiveSystems.pdf

\vspace{1mm}%\noindent
% \subsubsection{Task Planning Problem Generation}
\textbf{Task Planning Problem Generation (\emph{S1}):} 
Step \emph{S1} extracts the relevant requirements (\emph{S1a}) for creating the domain $\mathbb{D}$ %task plan 
and the numeric planning problem $\mathbf{P}$ (\emph{S1b}), cf. Section~\ref{sec:numericPlanning}. 
The relevant requirements from Table~\ref{table:missionReqs} %($\Phi_{1\text{-}4}$) 
are formalised to inform the execution of feasible actions $\mathcal{A}ct$ for the CPHS agents.
% as part of the planning actions $\mathcal{A}ct$ (see Sec.~\ref{sec:background}). 
Concretely, constraints $C_1$--$C_3$ apply to action \emph{Move},
\begin{multline}
 \forall a: \mathcal{A};\ l_i,l_j: \mathcal{L}\ | 
\ \mathit{Path}(l_i,l_j) \land 
\mathit{AgentLoc}(a)=l_i\ \land 
\\
\mathit{Empty}(l_j) 
\implies\ \textit{Move}(a,l_i,l_j)   
\end{multline}
\noindent
resulting in move actions that are available only when there is a path to the target location ($C_1$), the agent is at the initial location ($C_2$), given by $\mathit{AgentLoc}:\mathcal{A}\to \mathcal{L}$,
and the target location is empty ($C_3$), given by Boolean $Empty: L \mapsto \mathbb{B}$.

For the \emph{Do} action, only constraint $C_4$ applies, yielding
% % Constraint $C_4$ applies to the action \emph{Do}. 
% We use the agent's probability of successfully completing a task and the threshold~$\gamma$ to create a (non-probabilistic) Boolean expression, 
% $\forall a: A;\ l: L\ |\ Psucc(a)\geq \gamma\ \bullet\ \text{Do}(t,l)$.
% We added three additional constraints to the dotask action: agents must be at the task's location to perform it, and the task must be available and not yet completed at that location, s.t.,
\begin{multline}
 \forall a: \mathcal{A};\ l: L;\ t: T\ |
 AgentLoc(a)=l \land TaskLoc(t)=l \land 
 \\
 \neg TaskDone(t) \land
 \ Psucc(a,t)\geq \gamma\ \implies\ \emph{Do}(a,t,l)
\end{multline}
resulting in do actions executable when an agent is at the task's location, the task is available and not yet done at that location, and provided that the agent's competency exceeds threshold~$\gamma$.

The numeric planning problem is written in the PDDL language~\cite{younes2004ppddl1}. Figure~\ref{fig:inputJsonFile}b shows the PDDL domain $\mathbb{D}$, where the non-highlighted parts are generic and reusable. 
The domain comprises three types $\mathbb{T}$: locations, tasks and agents. Predicates~$\mathbb{P}$ are defined for the agent's location (\emph{agent\_at}), paths (\emph{path}), empty locations (\emph{empty}), task locations (\emph{task\_loc}), and the completion of tasks (\emph{task\_done}). 
The two functions $\mathbb{F}$ specify the probability of an agent completing a task successfully (\emph{p\_success}) and track the travelling cost (\emph{travel\_dist}).  

Actions $\mathbb{A}$ exist to move an agent and perform a task. The \emph{Move} action precondition (Figure~\ref{fig:inputJsonFile}b, lines 13-15) requires an existing path, an agent at the initial location, and the next location to be empty. 
The effect of this action (lines 16-19) is that the agent moves to the specified location, the original location becomes empty, the destination location is no longer empty and the total travel time increases (line 19).
% . This increment (line 19) comes from the $distance$ value in the input file. 
The \emph{Do} action precondition (Figure~\ref{fig:inputJsonFile}b, lines 22-24) specifies that an agent must be present at the task location, the task must be in an undone state, and the agent must have a probability of successfully completing the task $\geq \gamma$. \changed{The function \emph{p\_success} associates an agent to its probability of task completion. In our example, as $\gamma=0.5$, only robots with at least 50\% chance of completing the task are allowed to take action \emph{Do}}. The effect of this action (line 25) results in the task being done.

Figure~\ref{fig:inputJsonFile}c shows an example of the planning problem $\mathbf{P}$ in PDDL. 
\changed{As before, non-highlighted parts are generic and reusable}
%signify \changed{its} generic \changed{p}arts. 
The problem-typed objects $\mathbf{O}$ are defined by the location, agent and task identifiers from the input file. 
The initial state $s_{init}$ (lines 6-14) comprises the travel cost initialised to zero (line 7); all existing paths defined by the symmetric relation of paths defined in the input file (line 8); the empty locations comprising all locations except the initial location of agents (line 9); the task locations (line 10) defined from task instances in the input file; the agents initial locations (line 11); and the probabilities of each agent succeeding with the tasks capable of performing (lines 12-14). 
When an agent cannot perform a task, this is initialised to zero (line 14). %These values act as conditional statements in the precondition of \emph{Do}. 
\changed{F}inally, goal $g$ entails that all tasks are done, i.e., constraint $C_5$, (line~16) is met and the optimisation objective of minimising travel distance (line~17) is achieved.

\vspace{0.5mm}%\noindent 
\emph{Example 2 (Task Planning Model).}  
We defined the task planning model in Figures~\ref{fig:inputJsonFile}b and \ref{fig:inputJsonFile}c, showing the domain $\mathbb{D}$ and problem $\mathbf{P}$ in PDDL for our vineyard case study. 
The highlighted parts follow the colour palette of the corresponding parts in the input file from which these were obtained. %For example \textcolor{red}{...}

\vspace{1mm}%\noindent
% \subsubsection{Task Planning Problem Generation}
\textbf{Task Plan Generation (\emph{S2}):}\label{subsubsec:Pgen}
% \vspace{2mm}
% \noindent
% \subsubsection{Task Plan Generation}\label{subsubsec:Pgen}: 
Given the task planning formulation, defined in PDDL by a domain $\mathbb{D}$ and a problem $\mathbf{P}$, step \emph{S2} generates a feasible but uncertainty-agnostic task plan using an off-the-shelf numeric planner like ENHSP~\cite{scala2016interval}. 
The outcome of the planning problem is a plan $\Pi$, comprising a sequence of actions that transitions the CPHS from one state to another, ultimately achieving the goal state ($g$). 
% formally defined as a sequence of actions $\Pi=\langle a_1,a_2,...,a_n \rangle$, where each action $a_i\in \text{instance of } \mathbb{A}$, transforms the system from one state to another, ultimately achieving the goal state $g$. 

\vspace{0.5mm}%\noindent  
\emph{Example 3 (Task plan).} 
For our case study, an optimal plan that minimises the travelling cost is the following sequence of robot ($r1$) and worker ($w2$) actions: 
{\small \changed{$\langle$$Move$}($w2, l_1, l_4$), \changed{$Do$}($w2, t1l4, l_4$), $Do$($w2, t3l4, l_4$), $Move$($w2, l_4, l_7$), $Do$($w2, t1l7, l_7$),  $Do$($w2, t3l_7, l_7$), $Move$($r1, l_1, l_4$), $Move$($r1, l_4, l_5$), $Do$($r1, t2l5, l_5$), $Move$( $w2, l_7, l8$), $Move$($w2, l_8, l_9$), $Move$($r1, l_5, l_8$), $Do$($r1,$ $t2l8a, l_8$), $Do$($r1, t2l8b, l_8$),  $Do$($w2, t3l9, l_9$), $Move$($w2, l_9, l_6$), $Do$($w2, t1l6b, l_6$), $Do$($w2, t1l6a, l_6$)$\rangle$}. %This sequential plan is further discussed in the evaluation section. %Figure~\ref{fig:taskplan} shows this plan separating the actions scheduled for each agent.
% \footnote{A simple algorithm is used to divide the sequence plan by agents. Due to the constraint $\Phi_3$, tasks are added only when locations are empty. For example, location $l4$ is occupied by worker $worker1$ during the first three actions, therefore $r1$ moves after this.}

\vspace{1mm}%\noindent
% \subsubsection{Uncertainty Augmentation and Task Plan Refinement} 
\textbf{Uncertainty Augmentation and Task Plan Refinement (\emph{S3}):}
Step \emph{S3a} extracts uncertainty information from the \changed{p}roblem specification to generate the relevant probabilistic requirements and a probabilistic model of the plan in steps \emph{S3b} and \emph{S3c}, respectively. 
For the probabilistic model, it extracts the probability of task success ($PSuccess$), costs ($distance$ and $CostT$) and allowed number of task retries ($Retry$). \changed{F}or the probabilistic requirements, it extracts the minimum probability of \changed{mission} success $p_{succ}$.

Next, step \emph{S3b} formalises requirements $C_6$ and $O_1, O_2$ in PCTL. 
Thus, it uses information about the task plan generated in step \emph{S2} to define the final states: ``success", where all agents completed their allocated tasks successfully, and ``done", where no more actions are possible (due to mission success or failure). 
% as tasks were completed or retries their maximum number of retries. 
The PCTL requirements yielded are:
\changed{
%$C_6$: $\mathcal{P}$=?[$\mathcal{F}$ ``success"]$\geq p_{succ}$;
$C_6$: $\mathcal{P}_{\geq p_{succ}}$[$\mathcal{F}$ ``success"];
%$O_1$: minimise the total expected cost $\mathcal{R}$=?[$\mathcal{F}$ ``done"]; and
$O_1$: minimise the total expected cost ($\emph{min}\;\mathcal{R}_{=?}^{cost}$[$\mathcal{F}$ ``done"]); and
$O_2$: maximise the probability of mission success ($\emph{max}\; \mathcal{P}_{=?}$[$\mathcal{F}$ ``success"])}.

In step \emph{S3c}, a parametric DTMC model $\mathcal{D}$ of the plan is created, enhanced with uncertainty information. 
To create this model, we extract the number of actions ($n_{act}$) assigned to an agent in the task plan from step \emph{S2} and the number of tasks assigned to an agent ($n_{t}$)  that allow for retries. 
Model~$\mathcal{D}$ contains state variables for each agent in the task plan, defined by the tuple $s=(c, \overline{x}, )$, where $c\in[0,n_{act}+1]$ tracks the agent's actions increasing by one when an action is completed ($c=$0 to $n_{act}$), or when a task has failed ($c = n_{act}+1$); and $\overline{x}=x_1,x_2,...$ is a collection of variables such that $x_i$ tracks the attempted retries of a task. 
The upper bound of variable $x_i$, i.e., $\hat{x_i}$, is a parameter of $\mathcal{D}$ % $x_i\in[0,param]$. 
% Its 
and its value is synthesised in step \emph{S4} to yield a robust plan. %ynthesised as explained later, defines the number of times a retry is allowed. 
The maximum number of retries cannot exceed the limit defined in the problem specification (\emph{step S0}),
% by the user in the input file, 
hence $\hat{x_i}\in[1, Retry(t)]$. 
The total number of state variables is the sum of  agents selected to undertake tasks in the CPHS and the set of tasks with retries per agent. 

% $\sum_{a} n_{t_a} + |\text{agents in task plan}|$, where $n_{t_a}$ is the number of tasks with retries per agent.

Transitions in model $\mathcal{D}$ per agent $a$ are defined based on the sequence of actions it must perform. Move actions increase $c$ by one. 
Attempting a task $t$ with $Retry(t)=0$ results in $c\mathrel{+=}1$ with probability $PSuccess(a,t)$ representing a succeeded attempt, and in $c=n_{act}+1$ with probability 1-$PSuccess(a,t)$ when failed. 
Finally, attempting a task $t$ with $Retry(t)>0$ has two possible transitions depending on the value of $\overline{x}$. Let $x\in\overline{x}$ be the state variable tracking the number of retries for task $t$,
\begin{itemize}
    \item if $x<\hat{x}$, then with probability $PSuccess(a,t)$ succeeds and $c\mathrel{+=}1$, and with probability 1-$PSuccess(a,t)$ fails increasing the number of retries $x\mathrel{+=}1$.
    % \item if $x<\hat{x}$, then with probability $PSuccess(a,t)$ succeeds and $c\mathrel{+=}1$, and with probability 1-$PSuccess(a,t)$ fails,  $c=n_{act}+1$ and the number of retries $x\mathrel{+=}1$.
    \item else, it fails resulting in $c=n_{act}+1$.
\end{itemize}

Rewards are defined over transitions so that a \emph{Move} action incurs a travelling cost $distance$, and \emph{Do} action a cost of $CostT(a,t)$ per agent-task pair $(a,t)$. 
This step yields the model $\mathcal{D}$ and PCTL properties in the EvoChecker language~\cite{gerasimou2015search}, used for \changed{the} task retries' synthesis in step \emph{S4}. 
% We explain its syntax with the help of the following example.

% \begin{figure}[t]
% \centering
%   \includegraphics[width=\linewidth]{Figures/evomodel.pdf}
%   \caption{Parametric DTMC and properties in EvoChecker.}
%   \label{fig:evomodel}
% \end{figure}

\vspace{0.5mm}%\noindent 
\emph{Example \changed{4} (Probabilistic Model).} 
Figure~\ref{fig:inputJsonFile}d shows the task plan augmented with uncertainty data. Lines 2-3 define the retry parameters and their range \changed{for} the meta-heuristic search. For instance, $w2$ assigned with task $t1l4$ which can retry up to five times is shown in line 2. Lines 4-5 capture the probability of success for each task assigned to each agent. Lines 6-7 define two values of state variable $c$ for each agent, \changed{named} as Final (i.e., robot succeeds with its tasks) and Fail. Each agent's behaviour is modelled as a separate module. Lines 8-22 describe worker \textit{w2} actions. Lines~9-12 initialise the state variables $c$ (line~9) and $\overline{x}$ (lines 10-12). The transition at line~13 shows a move action. Transition at lines 14-18 shows the attempt to complete task $t1l4$. Transition at lines~19-22 shows $w2$ failing to complete the task after the allowed number of retries. The reward structure (lines 25-30) shows the transition rewards with costs specified in the input file. The formalised properties in EvoChecker are shown in the white rectangle, with the minimum success probability value ($0.95$) extracted from the problem specification in step \emph{S0}.

\vspace{1mm}%\noindent
\textbf{Synthesis and verification (\emph{S4}):}
\label{subsubsec:Synthesis&ver}
To generate a Pareto-optimal set of verified plan solutions, our approach uses meta-heuristic search to search for combinations of task retries using genetic algorithms (GAs)~\cite{coello2006evolutionary}. Chromosomes represent the number of retries for each task. Internally, at the GA evaluation stage, our approach generates concrete DTMC models of the task plan with the parameters $\hat{x}$ fixed, %set to this number of retries. 
and employs PMC to analyse the set of extracted PCTL properties (bottom right of Figure~\ref{fig:inputJsonFile}d).
% previously introduced in the PRISM language. 
The Pareto-optimal front PF contains the values of the task plan's expected cost and success probability for objectives \changed{$O_1$} and \changed{$O_2$}, while the Pareto-optimal solutions set PS contains the values of the number of retries per task. 
% The output of step S4 is a set of verified plans $PS=\{\pi_1,\pi_2,...\}$.

\vspace{0.5mm}%\noindent 
\emph{Example \changed{5} (Pareto Front).} 
Figure~\ref{fig:inputJsonFile}e
shows the Pareto-optimal front found with 25 different combinations of decision points clustered into five regions. \changed{A single solution depicted in red shows the task retries allowed as a dictionary. The GA was set to 200 evaluations with a population size of 25.} %Population

% \begin{figure}[t]
% \centering
%   \includegraphics[width=0.7\linewidth]{Figures/pareto.pdf}
%   \caption{Pareto-optimal set of solutions.}
%   \label{fig:pareto}
% \vspace{-4mm}
% \end{figure}

\vspace{1mm}%\noindent
\textbf{Incremental Adaptation (\emph{A1--A3}):}
\label{subsubsec:IncrementalAdaptation}
%Our approach generates an initial Pareto-optimal set of verified plans at design time; the current plan is then adjusted in response to changes at runtime. %\footnote{For example, if the user prefer a trade-off in the optimisation objective values, they would prefer knee points in the Pareto front; extreme points are preferred when an optimisation objective has a preference over the other.} 
%
%failures in the tasks and changes in the problem specification. For the number of failures, it tracks how many times each task has failed (C1). For the problem specification, it checks when a new problem $\mathcal{M'}\neq \mathcal{M}$ has a change in\footnote{This list is an example, but it can be extended, for example, capturing changes in number of retries $Retry(a,t)$.}:
%
At runtime, the implemented plan is adjusted in response to changes, leveraging the self-management and adaptation capabilities for CPHS offered by the MAPE-K loop~\cite{kephart2003vision}.
% We explain the runtime activities with the help of the MAPE-K loop~\cite{kephart2003vision}. The MAPE-K loop is a process model used in autonomic computing, consisting of four key phases—Monitor, Analyse, Plan, and Execute—supported by a Knowledge component to enable self-management and adaptation of systems. 
The Monitor checks for the following changes and disruptions:
\begin{itemize}
    \item C1: a task failure; 
    \item C2: a change in the minimum success  probability $p_{succ}$;
    \item C3: a change in the minimum assignment probability $\gamma$;
    %\item C3: number of retries $Retry(a,t)$ for an agent $a$ and task $t$;
    \item C4: a change in an agent's probability $PSuccess(a,t)$.
    % \footnote{For a task $t$, where these probabilities may vary dynamically at runtime, based on sampling and Bayesian updating techniques~\cite{camilli2022taming}.}
\end{itemize}
%The monitor phase also creates a plan $\pi_t$ up to the current time $t$ with the actions completed by the CPHS task planning. This plan has actions: travel, a task done and a task retry, and it is used by the Analyse phase as follows.
%
The Analyse phase receives the change type C$\in$\{C1,C2,C3,C4\} and analyses the current state of the system compared to its status in the knowledge base. 
Based on the analysis result, a strategy is selected by the Plan phase. 

The adaptation process is shown in Algorithm~\ref{alg:analysis-adapt}. 
Given a change C, a deployed plan $\Pi$, the time of the change $\tau$ and an initial set of verified and robust plans $PS$ (i.e., the Pareto front set), our approach results in an adaptation related to different parts of the task planning approach (A1, A2 or A3) as shown in Figure~\ref{fig:overview}. 
For all changes, our algorithm follows a similar data flow: (a)~reduce the $PS$ set by removing incorrect plans that do not comply with the change, 
(b)~if our current plan is part of this $PS$, continue without adaptation, 
(c)~otherwise, check if another plan exists ($PS.nonempty$) and deploy a new plan. 
If no plan exists, re-plan from step \emph{S0}.

Let $\Pi_j\in PS$ be a verified plan. Let $\Pi_j[k], k\in\mathbb{R}^{+0}$ denote the action (starting) at time $k$ in $\Pi_j$; $\Pi_j[:k]$ all actions up to time $k$ and $\Pi_j[k:]$ from $k$ onwards.
% \footnote{Here, we use the notion of a time unit as the duration of an action.} 
Here, we use the notion of a time unit as the duration of an action.
A change %is said to 
occurs at time \(\tau\) if it happens at time \(\tau\) or later, but before \(\tau + 1\). %The deployed plan $\Pi \in PS$ is a global variable.

When C1 occurs at time $\tau$ (lines 1-2), REDUCE\_PS\_TF($PS,\tau,t$) selects a new set of plans $PS'\subseteq PS$ with plans that comply with the current \changed{progress} ($\Pi_j[:\tau]=\Pi[:\tau]$) \changed{and has (a retry of) do task $t$ at the next step $\Pi_j[\tau+1]$}. % that allow for a retry of $t$ at the next step ($\Pi_j[\tau+1]!=$ retry of $t$). 
Similarly, when C2 occurs (lines 3-4), REDUCE\_PS\_PSUCC($PS, \tau, p_{succ}'$) selects a new set of plans $PS'\subseteq PS$ with plans that comply with the current \changed{progress} as in C1 and that have a \changed{mission} success probability greater than or equal to the new $p_{succ}'$.

When C3 occurs (lines 5-6), REDUCE\_PS\_PASSIGN($PS,\tau,t,\gamma'$) first checks if any (undone) task in $\Pi[\tau:]$ has $PSuccess(a,t)\leq\gamma'$. If this holds, it yields an empty $PS'$ triggering line~\ref{A3} in Algorithm 1. This change entails that as all $\Pi_j\in PS$ follow the same allocation from \emph{S2}, when the allocation is invalid all plans become invalid.
Finally, when C4 occurs, REDUCE\_PS\_PTASK($PS$,$t,a,p'$) first checks whether \changed{the new probability $p'$ of succeeding with $t$ by agent $a$} applies to any (undone) tasks in $\Pi[\tau:]$. If it does, the function proceeds similarly to C3, checking if the new probability $p' > \gamma$; if not, the process is reset from step \textit{S0}. If the condition holds, a new set $PS' \subseteq PS$ is generated from \emph{S3} for the tasks in $\Pi[\tau:]$, with the updated probabilistic information $p'$.

Adaptations A1-A3 occur at various stages of the task planning process, as illustrated in Fig.~\ref{fig:overview}. A minor adaptation A1 might be required by changes C1 or C2, requiring only plans from $PS$. A medium adaptation A2 might be required by C3 or C4 when plans follow the allocation constraints but new probabilistically verified plans must be obtained. A major adaptation A3 might be required by C3 or C4 when allocations are no longer valid as they violate constraints related to the allocation threshold $\gamma$.

% \textit{Theorem (Algorithm correctness). Algorithm~\ref{alg:analysis-adapt} terminates with $\Pi$ guaranteed to be correct, if such a plan is found.}. 

\noindent 
\textbf{Algorithm Correctness.}
Algorithm~\ref{alg:analysis-adapt} terminates for a finite $|PS|$ and without any loops.
A finite $|PS|$ is ensured as $PS$ results from the synthesis step \emph{S4} (see Fig.~\ref{fig:overview}), which is set to a finite number of evaluations. 
The correctness of the solutions produced by the adaptation algorithm is ensured through the following guarantees: (a)~changes C1 and C2 only modify the plan by selecting from previously generated correct plans in $PS$; (b)~change C3 solely impacts the allocation process (as the $\gamma$ threshold only impacts the task allocation), hence adaptations result in a new task planning problem from \emph{S0}, with the new problem specification guarantees a new correct set of solutions; (c)~change C4 checks compliance with both planning and probabilistic constraints. This change can proceed as C3, or generate new plans from \emph{S3}. As the input plan in \emph{S3} guarantees compliance with the planning constraints, from \emph{S3}, a new set of verified plans (in \emph{S4}) is then guaranteed to be correct.

Although we can guarantee the correctness of the generated plans through our adaptation process, we cannot guarantee that a plan exists, for example, when too strict requirements are set (e.g., minimum probability of success $>$0.99) or the planning problem is unsolvable (e.g., when no path exists to reach one of the tasks). Moreover, to reduce complexity, we limit our approach to only one change per time unit for this paper. This assumption holds for agents working in parallel. A plan \changed{$\Pi$} contains the actions of all agents, while an individual agent's plan contains only that agent's actions. The key difference is that multiple actions can be executed simultaneously. By restricting ourselves to one change per time unit, adaptations impact the plans of all agents at once. Future work will investigate ways to relax this constraint.

% \vspace{-2mm}
\setlength{\textfloatsep}{0pt}% Remove \textfloatsep
\begin{algorithm}[t]
\caption{Task plan adaptation algorithm}\label{alg:analysis-adapt}
\begin{algorithmic}[1]  
\small
\Require Monitored change C 
\Require  Global var.  $\Pi$ \Comment{current plan}
\Require  Global var.  $\tau$ \Comment{time of change}
\Require Global var. $PS$ \Comment{set of verified plans}
\Require Global var. $M$ \Comment{Definition~\ref{def:probSpec} (Problem specification)}
\If{C=C1} \Comment{C1: task $t$ failure}
	\State $PS'\leftarrow$ REDUCE\_PS\_TF($PS$,$\tau$,$t$)

\ElsIf{C$=$C2} \Comment{C2: \changed{mission success} $p_{succ}'$ change}
	\State $PS'\leftarrow$ REDUCE\_PS\_PSUCC($PS$,$\tau,p_{succ}'$)
\ElsIf{C=C3} \Comment{C3: prob. of assignment $\gamma'$ change}
        \State $PS'\leftarrow$ REDUCE\_PS\_PASSIGN($PS$,t,$\gamma'$)  \Comment{via PMC}
\ElsIf{C=C4} \Comment{C4: task $t$ success from $a$ changed to $p'$}
	\State $PS'\leftarrow$ REDUCE\_PS\_PTASK($PS$,$t,a,p'$)  \Comment{via PMC}
\EndIf
\If{($\Pi\in PS'$)}
    \State \Return \changed{$\Pi$} \Comment{No adaptation required (N/A)}\label{NA}
%\ElsIf{!($\Pi\in PS'$) $\&$ $(PS'.nonempty)$}
\ElsIf{\changed{$(PS' \neq \emptyset)$}}
%	\State $\Pi\leftarrow\Pi_j, \Pi_j\in PS$     \Comment{A1/A2}
	\State \changed{\Return SELECT\_NEW\_PLAN($PS'$)} \Comment{A1/A2}
%	\State $\Pi\leftarrow\Pi_j, \Pi_j\in PS$     \Comment{A1/A2}
\Else
	\State \changed{$PS \leftarrow$ GENERATE\_NEW\_PLANS($M, \Pi, \tau$) }
	%generate new plans from \emph{S0} 
	\Comment{A3} \label{A3}
		\State \changed{\Return SELECT\_NEW\_PLAN($PS$)}
\EndIf
 \State Return %\Comment{No adaptation required (N/A)}\label{NA-1}
\end{algorithmic}
%{\footnotesize*=new $PS$ only with undone tasks.}
\end{algorithm}

\input{Tables/table-RQ1}

% \vspace{-1mm}
\section{Evaluation}
\label{sec:evaluation}
\vspace{-1mm}
\subsection{Research questions}

% In this section, we evaluate the effectiveness and efficiency of our approach.

% \gricel{This can be divided into two: how much we reduce the size of the probabilistic model to be verified, and how close to optimal solutions we get}
% \subsection{Research Questions}

\noindent
\textbf{RQ1 (Effectiveness): How effective is our hybrid approach, in terms of solution quality and computational performance, when compared to an optimal solution derived from a full MDP model of the problem specification?} %\alexandros{How effective is our hybrid approach, in terms of solution quality and computational performance, when compared to the optimal Pareto fronts derived from a full MDP model of the problem specification?}
To mitigate the state explosion problem, our approach leverages heuristic numerical planners for task partitioning and scheduling, and verifies augmented plans using PMC. We compare the effectiveness of our verified plan solutions against those from a full MDP (optimal) baseline synthesised using PRISM~\cite{kwiatkowska2011prism}. % by modelling the task planning problem as a full MDP.

\noindent
\textbf{RQ2 (Adaptation): How effective is our approach in dealing with runtime changes by adapting the plan when necessary and reducing latency costs?} We assess the effectiveness of our runtime adaptation to changes C1-C4 and discuss reductions in replanning latency.% through our incremental adaptation.

%To evaluate the efficiency of our runtime adaptation approach, we inject the previously introduced changes C1-C4 to a pre-deployed task plan and describe the adaptation process selected to effectively recover from these changes. 

\noindent
\textbf{RQ3 (Efficiency): How computationally efficient is our hybrid approach when we increase the number of tasks and agents in the problem specification?} % \alexandros{How does the computational efficiency of our hybrid planning approach change as we increase the number of tasks and agents in the problem specification?}}
We investigate the efficiency of our approach when increasing the planning problem in the number of tasks and the number of agents.

\subsection{Experimental setup}
%AIRPLANE
%\textcolor{green}{
We assess the effectiveness, adaptability, and efficiency of our hybrid approach and compare it against a full MDP baseline derived from the problem specification. 
We use several problem instances of varying complexity and numbers of tasks and agents, including those (M1, M2, M3, M4) from Table~\ref{table:fullMDP}.
% discussed in Section \ref{sec:results}.

\textbf{RQ1 (Effectiveness)}. To evaluate the effectiveness of our approach compared to an optimal baseline, we generated a full MDP from the problem specification in Section~\ref{sec:ProblemFormulation}.
By obtaining policies from this full MDP, we compute optimal Pareto fronts that serve as baselines for comparison. %These optimal fronts serve as a benchmark against which we evaluate the solutions produced by our hybrid method.
We apply our approach to several problem instances of increasing complexity %(as defined in Section~\ref{sec:ProblemFormulation}) 
and measure solution quality, execution times (mean and standard deviation), and the number of solutions \changed{over 30 runs}. We also compare the model sizes of the MDP, and the DTMC generated in step ~\emph{S4} with parameters set to the maximum number of retries.

\textbf{RQ2 (Adaptation).} To investigate how our approach manages runtime changes (e.g., task failures, probability value changes), 
%To illustrate the adaptation process, 
similar to~\cite{gerasimou2014efficient}, we present a scenario where all changes (C1-C4) and adaptations (A1-A3) happen during a typical execution of our industrial vineyard case study. We assessed latency reduction when applying different adaptation strategies at different steps of our hybrid planning approach.

\textbf{RQ3 (Efficiency)}. %Our final research question 
We examine how the computational efficiency of our approach changes as the problem size grows. 
Thus, we vary the number of tasks (10 to 13) and agents (2, 4, and 6) from our vineyard case study. Tasks were assigned random locations, while agents were deployed at location $l_1$. We report metrics as those for RQ1; however, given the high variation in the execution time results, we reported the geometric median following standard statistical guidance~\cite{fleming1986statistics}. 
%\textbf{RQ3}. We vary the number of tasks and agents  from our vineyard case study.}  We generate a testbed varying the number of tasks (10 to 13) and agents (2, 4 and 6) from our vineyard case study. Tasks were assigned random locations, while agents were deployed from $l0$. We generate 30 random scenarios for each configuration and report metrics as for RQ1. \textcolor{red}{Given the high variation in the execution time results, we reported the geometric average following standard statistical guidance}~\cite{}. As we were interested in efficiency, we set the GA to a larger GA population size (100) compared to RQ1.

\textbf{Implementation details}.
All experiments are run on an Intel Core i5 @2.40 GHz machine with 8 GB RAM, under Ubuntu 22.04.5 LTS 64-bit. Our hybrid approach is fully automated. For step \emph{S2}, we use a model-to-model (M2M) transformation implemented in Python to generate the PDDL files. We produce a task plan using the ENHSP numerical solver in the Unified Planning library \cite{unified_planning_UP}. A second M2M transformation generates the EvoChecker input files. %(.props for properties and a .dtmc for the pDTMC). 
Multi-objective optimisation at steps \emph{S3}-\emph{S4} is performed using EvoChecker~\cite{gerasimou2015search,gerasimou2018synthesis}, which orchestrates PRISM and the JMetal \cite{durillo2011jmetal} search framework with NSGA-II configured to 150 evaluations and a population size of 30 following guidelines~\cite{deb2002fast}. In RQ3, as we were interested in efficiency, we increased the population size to 100. Experiments use PRISM set to 8~GB memory limit. 
%
% The runtime adaptation algorithm, Algorithm~\ref{alg:analysis-adapt}, is implemented in Java. 
Our open-source code is available at~\cite{approachGithub}.

%We provided a testbed with variations of our case study to test the feasibility and scalability of the synthesis of verified task plans. 
%Finally, for the runtime adaptation, we used our vineyard case study to demonstrate the feasibility of the Algorithm~\ref{alg:analysis-adapt}. 
%\url{https://github.com/Gricel-lee/EfficientPlanAdaptation}

\begin{figure*}[htbp]
    \centering
    \begin{minipage}{0.24\textwidth}  % First figure (a)
        \centering
        \includegraphics[width=\linewidth]{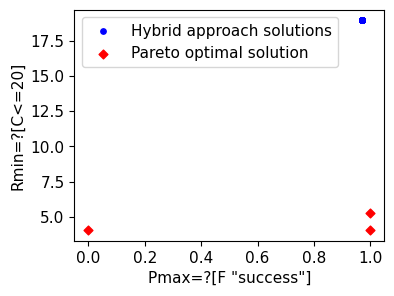}  % Replace with your figure file
        % \caption*{(a) Problem $\mathcal{M}_1$}
        \label{fig:RQ1-M1a}
    \end{minipage}\hfill
    \begin{minipage}{0.24\textwidth}  % Second figure (b)
        \centering
        \includegraphics[width=\linewidth]{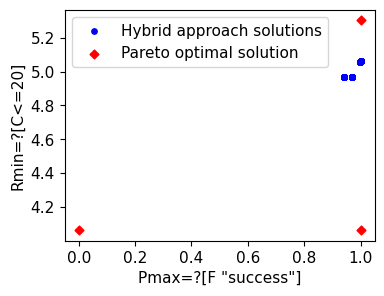}  % Replace with your figure file
        % \caption*{(b) Problem $\mathcal{M}_2$}
        \label{RQ1-M2b}
    \end{minipage}\hfill
    \begin{minipage}{0.24\textwidth}  % Third figure (c)
        \centering
        \includegraphics[width=\linewidth]{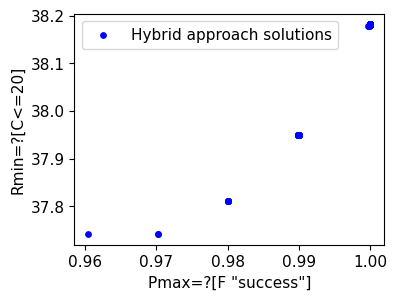}  % Replace with your figure file
        % \caption*{(c) Problem $\mathcal{M}_3$}
        \label{RQ1-M2c}
    \end{minipage}\hfill
    \begin{minipage}{0.24\textwidth}  % Fourth figure (d)
        \centering
        \includegraphics[width=\linewidth]{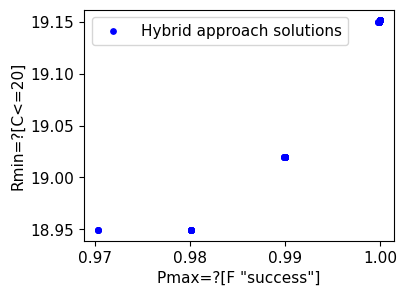}  % Replace with your figure file
        % \caption*{(d) Problem $\mathcal{M}_4$}
        \label{RQ1-M2d}
    \end{minipage}
    \vspace{-7mm}
    \caption{Pareto front from optimal baseline and our hybrid solution for problem instances. \changed{From left to right:} $\mathcal{M}_1, \mathcal{M}_2, \mathcal{M}_3, \mathcal{M}_4$.}
    \label{fig:ParetoFronts-fullMDP}
    \vspace{-2mm}
\end{figure*}

\subsection{Results and Discussion}
\label{sec:results}
\noindent
\textbf{RQ1 (Effectiveness).} %Since probabilistic model checkers PRISM nor Storm~\cite{} were able to solve the multi-objective verification problem outlined in Section~\ref{subsubsec:Synthesis&ver}---reachability reward properties are not supported for MDPs in multi-objective model checking---, we verified the bounded cumulative reward R$=?$[C$<=$20] and P=?[F ``success"] for the full MDP and our hybrid approach. We use PRISM to obtain the optimal Pareto front.
To generate the optimal Pareto front, we \changed{tested} the multi-objective verification capabilities of the widely-used PMC tools PRISM~\cite{kwiatkowska2011prism} and Storm~\cite{hensel2022storm}.
However, neither tool 
%two tools. However, neither PRISM nor Storm~\cite{hensel2022storm} were able to
could solve the multi-objective verification problem outlined in Section~\ref{subsubsec:Synthesis&ver}---reachability rewards for multi-objective properties are not supported.  Thus, we verified the bounded cumulative reward $\mathcal{R}_{min}$=?[$\mathcal{C}$$<=$20] and the reachability property $\mathcal{P}_{max}$=?[$\mathcal{F}$ ``success"] simultaneously for both our approach and PRISM/Storm using  the full MDP.

For the first two models (M1 and M2) we were able to generate Pareto solutions for both the full MDP and our approach. 
%both, the full MDP and our approach, Pareto solutions. 
However, due to an explosion in the size of the MDP models, we could not do the same for M3 and M4---M3 was parsed but could not be checked (Table~\ref{table:fullMDP}). Our hybrid approach successfully generated plans for all models.
Since we resolved the allocation and scheduling separately, the probabilistically checked models were reduced by several orders of magnitude. For example, M1 was reduced from 35,081 states to 14, and 172,714 transitions to 17. \changed{This also resulted in a lower execution time compared to the full MDP.} % using our hybrid approach.%  is also reduced , however, increases from seconds, to verify the full MDP, to minutes using our hybrid approach given the time-consuming evolution of the solution set using GA. 
Figures~\ref{fig:ParetoFronts-fullMDP}a and~\ref{fig:ParetoFronts-fullMDP}b show that our approach generates dominated solutions while the full MDP yields the optimal Pareto front. We continue evaluating each model's results separately.

For M1 we only allowed for one retry. Our approach first generated a plan to deploy only one robot. Therefore, we only find one single solution where the robot can retry all three tasks once. In comparison, the full MDP resulted in plans to deploy the robot or the human worker. These are the red diamonds close to $\mathcal{P}max$=?[$\mathcal{F}$ ``success"]=0.99. A human has a slightly higher probability of completion (0.99) than the robot (0.97) but incurs a higher expected cumulative cost.
% \textcolor{red}{Solution at $Pmax=?[F\ success]=0.0$ …}

For M2, our approach also deploys a single robot. However, as more retries are allowed, the Pareto front results in four solutions. Calculating the total number of solutions for this allocation results in $3^2$ (multiplying the number of retries per task), this is only 9 solutions to search by the GA---five dominated and four non-dominated. Models M1 and M2 are too small to benefit from our hybrid approach implementation resulting in unnecessary latencies. For these smaller problems, exhaustive searches can be considered instead.

For M3 and M4, we obtained several verified plans as shown in their Pareto front plots. Moreover, each of these Pareto front points corresponds to more than one combination of task retries. For example, for model M3 we obtained  \changed{8.51} Pareto front points on average \changed{(the solution in Fig.~\ref{fig:ParetoFronts-fullMDP}c shows these clustered near five points)}, while these correspond to a total of \changed{27.77} different verified plans on average. These results are the basis for our adaptation algorithm: we can change the plan to one with more retries, if needed, while ensuring plan correctness, even for complex models where full MDP verification is not feasible---though at the cost of optimality.

\begin{figure}[t]
    % \vspace{-2mm}
    \centering
    \includegraphics[height=0.2\textheight,width=0.93\linewidth]{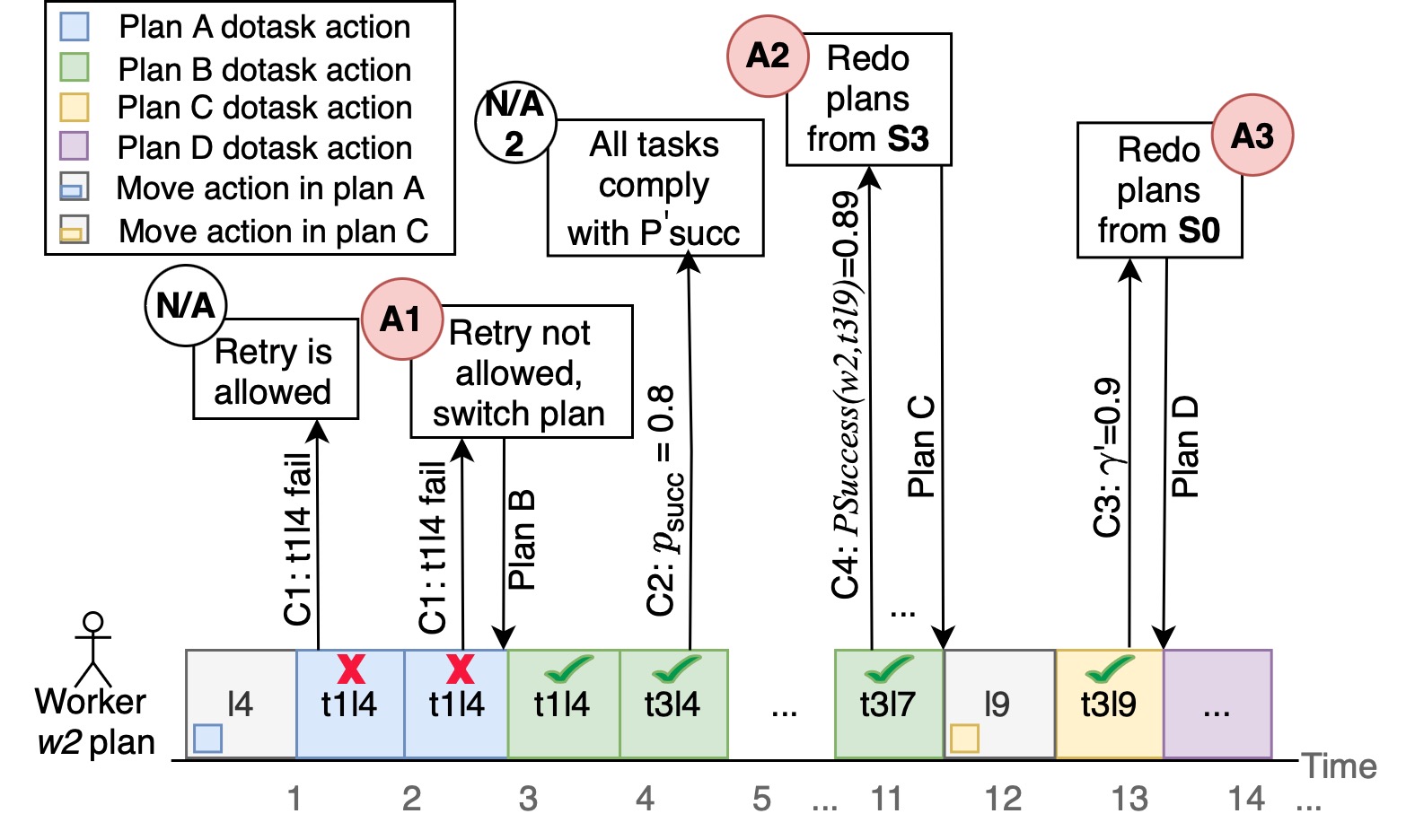}
    \vspace{-3mm}
    \caption{Sample scenario showing adaptation cases A1-A3.}%, signifying negligible, minor, and major disturbances, respectively.}
    % \vspace{-1mm}
    \label{fig:RQ2planWithAdaptations}
\end{figure}

\vspace*{1mm}\noindent
\textbf{RQ2 (Adaptation).}
%We want to check how Algorithm~1 works under different scenarios. For this, we implemented it as part of our tool where changes can be injected along the timespan. Given the finite number of monitored changes C1-C4 and adaptation scenarios A1-A3, we were able to exhaustively test each of these. 
Our adaptation algorithm accommodates a finite number of monitored changes C1-C4 and adaptation strategies A1-A3. Therefore, these were exhaustively tested in our provided tool~\cite{approachGithub}, where changes \changed{are} injected along the plan timespan. 
Here, we present the adaptation results on a continuous run for our industrial \changed{c}ase study showing all monitor changes and adaptations (see Fig.~\ref{fig:RQ2planWithAdaptations}). Worker~2 starts by travelling to $l4$.

%%Due to space limitations, we include further tests and the open-source code in our Github repository. 

%shows some of these results using our industrial case study after deploying worker $w2$. \textcolor{red}{As this exhaustively shows the adaptation strategies and monitored changes of our approach, }

%Plan 2
% \begin{itemize}
    % \item 
    \noindent$\bullet$ 
    \textbf{Case N/A:} At time 1 (i.e., between 1 and 2 time units as explained in Sec.~\ref{subsubsec:IncrementalAdaptation}), worker $w2$ attempts task \( t1l4 \) and fails. The task failure is already anticipated in the current plan, allowing for a $t1l4$ retry. 
% \end{itemize}

%Plan 24
%Solution:
%id: 24
%cost: 37.9482
%prob succ.: 0.9897
%Retries per task: [2, 1, 1, 2, 2, 5, 4, 6, 7, 2]
% \begin{itemize}
    % \item 
    \noindent$\bullet$ 
    \textbf{Case A1:} At time 2, task  \( t1l4 \) fails during a \changed{retry attempt. N}o second retry is allowed. Plan B, which allows for two $t1l4$ retries and is consistent with the current plan A progress, is found in the verified plans. After \changed{the second retry}, the task succeeds and proceeds with $t3l4$.

\noindent$\bullet$ 
\textbf{Case N/A 2:} At time 4, a change \changed{in the probability of mission success} $p_{succ}\!=\!0.8$ results in no \changed{adaptation}, as all \changed{plans already comply with such constraint update}.
% are above this threshold.

 % \item 
 \noindent$\bullet$ 
 \textbf{Case A2:} At time 11, the worker's probability $PSuccess(w2,t3l9)$ changes to 0.89 as tiredness start building up. This modifies the probability of plan success and a new verified plan is synthesised at \changed{\emph{S3}} for the remaining actions for both $w2$ and $r1$ agents \changed{(due to space limitations, we show only $w2$)}. A new plan C is deployed, which follows the same task and travel sequence \changed{---up to this point---} as Plans A and B.
 
 % \item 
 \noindent$\bullet$ 
 \textbf{Case A3:} At time 13, a change $\gamma'=0.9$ invalidates all verified plans obtained from \emph{S2} , as $w2$ can \changed{succeed} with $t3l9$ \changed{only with $0.89$} probability. Hence, a new plan is generated from \emph{S0} and the completion of tasks continues.
% \end{itemize}

This representative scenario shows how the planning adaptation transpires at runtime to avoid a task failure or the violation of a requirement. In RQ1, we discuss the computational costs of the different stages of the generation of plans. Here, we show that some of these costs can be avoided when no adaptation, or adaptations A1 and A2, are performed. At times 1 and 2, no replanning costs were incurred. At time 11, the execution time for numerical planning was avoided. %(for RQ1 models, this is measured in seconds).
 Meanwhile, the execution of the hybrid planning was only required at time 12, 
 %(for RQ1 models, this is measured in minutes), 
 emphasizing the efficiency gains achieved via the adaptation of hybrid planning techniques. The benefit of this incremental adaptation shows that replanning from scratch is not always necessary.
 In fact, task planning adaptation can leverage results from different stages of the planning process (such as the structure of a plan and the expanded set of verified plans), provided that the stages generating these results can guarantee their correctness after the change is applied.

\begin{figure}[t]
    \vspace{-2mm}
    \centering
    \includegraphics[width=\linewidth]{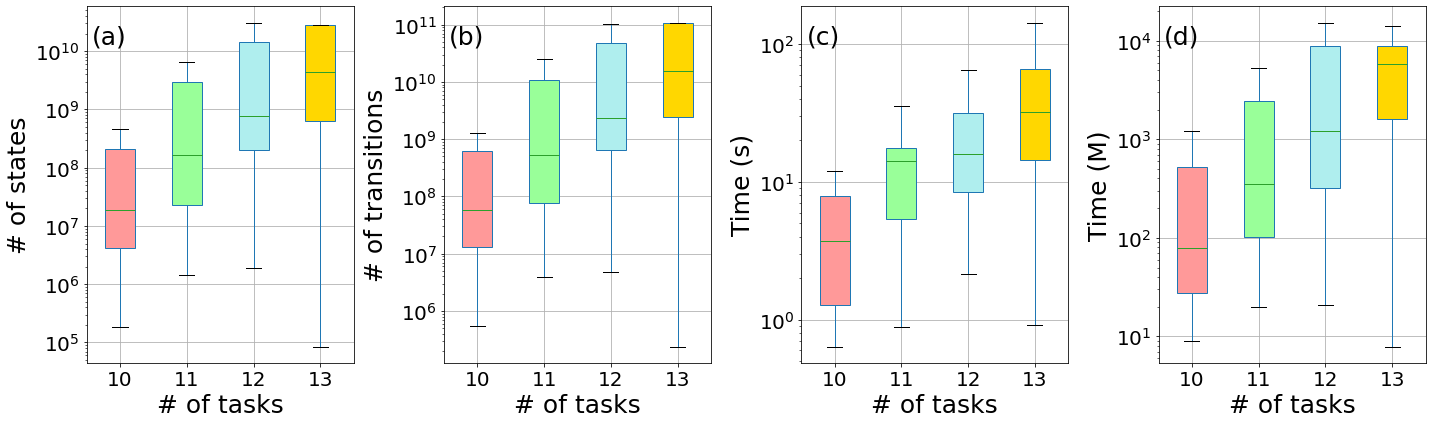}
    \vspace{-6mm}
    \caption{a) DTMC states, b) DTMC transitions, c) Numerical planner time, and d) Verif. \& search time for 10-13 tasks.}
    % \vspace{-2mm}
    \label{fig:RQ3}
\end{figure}

\noindent
\textbf{RQ3 (Efficiency)}. We conducted experiments for different combinations of the number of tasks and robots, with the results presented in Table~\ref{table:RQ2} and Figure~\ref{fig:RQ3}. The planner's computation time increases consistently for increments in both of these parameters, demonstrating relatively low variation in comparison to the verification and search counterpart. This is expected as verification and search perform multiple GA evaluations of the possible feasible plans. % internally using a (usually) expensive verification through PMC, which suffers from the state-explosion problem. In contrast, the numerical planner only generates a single non-probabilistic task plan. 
The verification and search times show significant variability depending on the number of tasks---in particular, on the number and locations of the tasks (see Fig.~\ref{fig:RQ3}c and d). As we randomly generated the task locations, these resulted in multiple planning problems, some easier to solve than others. This diversity explains the consistent increase in reported mean, but also in the standard deviation (SD).
\changed{For instance,} for 13 tasks, we have a variation in model size of $\pm 10.34 \times 10^{11}$ 
and $\pm38.20 \times 10^{11}$ geometric SD for states and transitions, respectively. The overall time required for verification and search increases as the total number of tasks grows, driven by the corresponding increase in plan complexity. The variations in the number of states and transitions follow a similar trend (Fig.~\ref{fig:RQ3}a and b). This determines the complexity of the probabilistic model, which increases on average with additional tasks.

For agent variations, all agents starting from \changed{$l_1$} resulted in a single plan per robot count. This results in a single probabilistic model size reported with retry parameter values set to the maximum number of retries. Increasing the number of agents increases the allocation complexity which, in turn, increases the heuristic planners' time. This increase is expected, as a greater number of agents introduces additional decision variables and constraints into the planning process. However, the verification and search times do not exhibit a discernible trend, highlighting their dependence on the sequential plan generated by the numerical planner. For instance, 2-agent and 6-agent cases yielded similar states and transitions, as only 2 agents (1 human, 1 robot) were actively assigned tasks in both scenarios. These findings indicate that the complexity of the verification part is driven by the number of robots in the numerical planning solution rather than by the initial number of agents. Furthermore, increasing either the number of tasks or agents leads to a significant rise in computational time.

\begin{table}[t]
\caption{Computational times and model sizes for RQ3.}
\vspace{-4mm}
\label{table:RQ2}
\sffamily
\begin{scriptsize}
\setlength{\tabcolsep}{2pt} % Adjust horizontal padding between columns
\renewcommand{\arraystretch}{1.1} % Adjust vertical padding
\begin{center}
\resizebox{0.49\textwidth}{!}{ % Resizes the table to fit the text width
\begin{tabular}{p{0.6cm} p{0.7cm}| p{1.7cm} p{2.2cm} p{1.9cm} p{1.9cm}}
\toprule
\begin{tabular}[c]{@{}l@{}}\# \\ tasks\end{tabular} & 
\begin{tabular}[c]{@{}l@{}}\# \\ agents\end{tabular} & 
\begin{tabular}[c]{@{}l@{}}Num. planner\\ time (SD) [s]\end{tabular} & 
\begin{tabular}[c]{@{}l@{}}Verif. \& search\\ time (SD$\times 10^{4}$) [s]\end{tabular} & 
\begin{tabular}[c]{@{}l@{}} \# states $\times10^{7}$ \\(SD$\times10^{9}$) \end{tabular} & 
\begin{tabular}[c]{@{}l@{}}  \# transitions $\times10^{9}$ \\(SD$\times10^{10}$) \end{tabular} \\ \hline 
10 & 4 & 3.71 (11.85) & 4,760.4 (\(0.06\)) & 1.89 (\(0.43\)) & 0.06 (\(0.15\)) \\
11 & 4 & 14.11 (10.35) & 21,301.05 (\(16.2\)) & 18.64 (\(20.00\)) & 0.55 (\(5.17\)) \\
12 & 4 & 16.02 (35.12) & 73,001.82 (\(28.2\)) & 84.39 (\(547.00\)) & 2.62, (\(155.00\)) \\
13 & 4 & 32.43 (49.70) & 348,623.4 (\(28.8\)) & 684.40 (\(1034.00\)) & 24.09 (\(382.0\)) 
\\ \hline
\hline
\begin{tabular}[c]{@{}l@{}}\# \\ tasks\end{tabular} & 
\begin{tabular}[c]{@{}l@{}}\# \\ agents\end{tabular} & 
\begin{tabular}[c]{@{}l@{}}Num. planner\\ time (SD) [s]\end{tabular} & 
\begin{tabular}[c]{@{}l@{}}Verif. \& search\\ time (SD) [m]\end{tabular} & 
\begin{tabular}[c]{@{}l@{}} \# states (SD) \end{tabular} & 
\begin{tabular}[c]{@{}l@{}}  \# transitions (SD) \end{tabular} \\ \hline 

10 & 2 & 0.53 (0.04) & 7.54 (11.65) & 1,872,856 & 3,915,885 \\
10 & 4 & 2.10 (0.09) & 13.92 (2.38) & 1,825,976 & 3,822,145 \\
10 & 6 & 9.39 (0.19) & 15.31 (1.38) & 1,872,856 & 3,915,885 \\ 
%\hline
\bottomrule
\end{tabular}
}
\end{center}
\end{scriptsize}
% \vspace{-2mm}
\end{table}

\vspace{2mm}\noindent
% \subsection{Threats to validity} 
\textbf{Threats to validity. } 
% \vspace{-1mm}
We reduce \textbf{construct validity threats} \changed{d}ue to simplifications in the specifications of the planning problem and adaptation tactics selection by using an industrial case study from our ongoing European-project collaboration. This paper is aligned with their vision of long-term adaptation for their CPHS tasks. We mitigate threats in the adaptation cases of  \changed{Algorithm~\ref{alg:analysis-adapt}} by exhaustively testing each of the possible adaptation scenarios automatically through our tool.

We mitigate \textbf{external validity threats}  that could affect the generalisation of our approach by using off-the-shelf numerical planners and the search-based software engineering tool EvoChecker~\cite{gerasimou2015search}; the latter using the widely used PRISM model checker internally. This also mitigates introducing errors in the relevant search algorithms underpinning our approach. We use the widely used PDDL2.1~\cite{younes2004ppddl1} language to define our numerical planning problem, and the PRISM-based EvoChecker language for the multi-objective optimisation part. For consistency, we verified the full MDP in PRISM~\cite{kwiatkowska2011prism}.

%See: https://eprints.whiterose.ac.uk/178508/1/MDPPolicySynthesis_ASE2021.pdf

\section{Related work}
\label{sec:relatedWork}

\changed{Multiple hybrid approaches have been proposed to address the multi-agent task planning problem and variants (task allocation, task scheduling and motion planning~\cite{messing2022grstaps})~\cite{fang2022automated,ham2021human,chen2021integrated}. Reviews such as \cite{chakraa2023optimization} present an overview of some of these hybrid approaches. These include combining GAs with mixed integer linear programming~\cite{zhou2019balanced}, Branch and Bound (BnB)~\cite{martin2021multi}, Q-learning~\cite{alitappeh2022multi}, game-theory~\cite{martin2023multi}, clustering~\cite{saeedvand2019robust,janati2017multi} and simulated annealing~\cite{junwei2014study}. KANOA~\cite{vazquez2024scheduling,vazquez2022scheduling} combines constraint solving with GA and PMC for the allocation and scheduling of tasks. In contrast, our solution uses GA and PMC for the verification of probabilistic properties and task retry synthesis, while efficiently pre-solving the numerical task planning problem through heuristic methods. PMC has become a prominent technique for ensuring reliability under uncertainty \cite{kwiatkowska2010advances,calinescu2015decide}, successfully applied in task planning problems~\cite{camara2020coadaptation}. However, its limited adoption remains due to the state explosion problem, motivating our approach.} %, task planning in mobile service robots---and the adaptation of software architecture---is addressed through PMC. 

Prior work on task planning under uncertainty, emphasising cyber-physical systems, has been studied using various approaches~\cite{zhao2024bayesian,pan2022failure,ye2024quantitative}. 
Random resource availability is considered in \cite{sanchez2024automated}, while \cite{alirezazadeh2022dynamic} considers dynamic task allocation through replanning or task reallocation. For self-adaptive systems, frameworks such as ROSRV, \cite{huang2014ROSRV}, and runtime verification approaches \cite{ferrando2018runtime,zudaire2022assured} can check the safety and temporal properties during system execution. However, these and many similar approaches~\cite{sanchez2024automated} do not provide explicit adaptation strategies. 
\changed{Finally, for CPHS, different variants, including human-in-the-loop~\cite{fischer2021loop}, human-on-the-loop~\cite{li2020explanations,ham2021human}, and human-machine-interaction~\cite{cleland2022extending} for task planning, have been extensively studied. Specifications such as``\textit{humans can reject plan solutions}"~\cite{wang2022toward} will be explored in future work.}

\section{Conclusions and Future Work}
\label{sec:conclusions}
We presented a hybrid adaptive task planning approach to generate correct and verified CPHS plans. 
% By leveraging a combination of
By combining numerical planning methods and probabilistic model checking, our approach effectively decomposes task planning into deterministic and uncertainty-augmented stages. Integrating meta-heuristic search enables synthesising Pareto-optimal plans that manage uncertainties while meeting formal probabilistic requirements. 
%Through an industrial case study, we demonstrated significant improvements in scalability compared to full MDP policy synthesis.% an optimal solution. 
Our approach produced plans for larger problems that policy synthesis based on full MDP and verification via probabilistic model checking failed. 
% By identifying their relevant requirements, we show the incremental plan adaptation process, where adaptations happen along different stages of our hybrid planner. 
Incremental adaptation also yields computational time savings when possible. %Finally, we show the impact of adding more robots and tasks, 

% Future work will extend our approach to support multiple simultaneous changes and other requirements relevant to our CPHS problem, e.g., 
% human fatigue levels or sensor failures through real-time feedback from agents.
% We will also investigate using advanced numeric planners and constraint solvers~\cite{micheli2025unified}, and decentralised planning~\cite{calinescu2015decide}. 
% Finally, we will explore using graphical editors for mission specification~\cite{ye2024quantitative,predoaia2024tree}.
% Finally, improving the scalability of plan verification for larger mission sizes remains a key direction for further research.

Future work will extend our approach to support multiple simultaneous changes and other requirements relevant to our CPHS problem, e.g., human fatigue levels or sensor failures instrumented through real-time feedback from agents. 
We will also investigate using advanced numeric planners~\cite{micheli2025unified} and decentralised adaptation~\cite{calinescu2015decide}. Finally, we will explore using graphical editors for mission specification~\cite{predoaia2024tree}.

\vspace{1mm}
\changed{
\textbf{Acknowledgements.} This research was supported by the Europe Horizon projects AI4Work (101135990) and SOPRANO (101120990), and by the ULTIMATE project funded by the Advanced Research and Invention Agency, UK. We thank Alessandro Valentini, Elisa Tosello and Andrea Micheli from Fondazione Bruno Kessler for their helpful support on the numerical planner, and Quinta Do Castro and the University of Trás-os-Montes for their help with the case study.
}

\bibliographystyle{IEEEtran}
\bibliography{SEAMS}  % The .bib file

\end{document}

%% file: Tables/table-RQ1.tex
\begin{table*}[]
\caption{Hybrid approach compared to a full-MDP model}
\vspace{-4mm}
%\url{https://docs.google.com/spreadsheets/d/1Iwyae02-TFwfa7iDuavsNNzRzTQacH21bkYkkwidn8A/edit?gid=1195719558#gid=1195719558}}
\label{table:fullMDP}
%\centering
\sffamily
\begin{scriptsize}
\begin{center}
\resizebox{\textwidth}{!}{ % Resizes the table to fit the text width
\begin{tabular}{p{0.5cm}|p{0.1cm}p{0.1cm}p{0.4cm}p{1.85cm}|p{1.0cm}p{1.2cm}p{1.6cm}p{0.4cm}|p{0.8cm}p{0.80cm}llp{1.5cm}}
\toprule%{2-12}
%\cline{2-12}

 \multicolumn{5}{c|}{\textbf{Planning problem}} 
 
 & \multicolumn{4}{c|}{\textbf{Full MDP verification approach}} 
 
 & \multicolumn{5}{c}{\textbf{Hybrid approach}}
 
 \\
\cline{2-11}
%\midrule
\hline

ID & $|T|$ & $|L|$ & retries & agents

& \multicolumn{1}{l}{\#states} & \multicolumn{1}{l}{\#trans} & \begin{tabular}[c]{@{}l@{}}PMC\\execution\\mean time\\ (SD) [s]\end{tabular} & $|PS|$ &  \# states* & \# trans*
& \begin{tabular}[c]{@{}l@{}}Numerical\\planner\\mean time\\ (SD) [s] \end{tabular}
& \begin{tabular}[c]{@{}l@{}}Verif.\\\& search\\mean time\\ (SD) [s]\end{tabular} 
&  \begin{tabular}[c]{@{}l@{}} $|PS|$ \\ mean (SD) \end{tabular} \\
%\midrule
\hline

% Evo:  104.22 1.8599999999999999
% Heur planner 0.333 0.025 

% Evo:  105.72 5.94
% Heur planner 0.298 0.025 

% Evo:  884.22 138.06
% Heur planner 0.331 0.025 

% Evo:  128.16 10.319999999999999
% Heur planner 2.028 0.082 

$\mathcal{M}_1$ & 3 & 9 & 1 & 1 human, 1 robot & 35,081 & 172,714 & 196.69 (85.38)  & 3  & 14 & 17 & 0.29 (0.02) & 104.2 (1.85)  & 1 (0.00) \\ %p = 0.9999999999333901, 0.999946000729
%problem3 in data

\textbf{$\mathcal{M}_2$} & 2 & 6 & 3 & 1 human, 1 robot & 134,581 & 541,767 & 128.53 (53.93) & 3  &  36 & 48 & 0.33 (0.02) & 105.7 (5.94)  & 4 (0.00)  \\ % Result: [(0.9999729999730006, 4.063123524275327)]
%problem1 in data

$\mathcal{M}_3$ & 3 & 9 & CS & 2 human, 2 robot & 270,100,547 & 1,351,524,022 & Timeout  & - & 398 & 553 & 0.33 (0.02) & 884.2 (138.06) & 47.80 (7.87)  \\ %0.99999999793485
%problem4 in data
$\mathcal{M}_4$ & 10 & 9 & CS & 2 human, 2 robot & OOM & OOM &  - & - & 1,825,976 & 3,822,145 & 
2.02 (0.08) & 128.1 (10.31) & 39.93 (3.17) \\ %0.9999999963720143
\bottomrule
\end{tabular}
}
\end{center}
\vspace{-1mm}
\text{*} from DTMC with max. number of task retries per agent (i.e., largest possible model checked).\\
OOM = out of memory. CS = retries set as in Table~\ref{table:vineyard_costs_prob}. All tasks in $\mathcal{M}_{1-2}$ are type $t3$. Model $\mathcal{M}_4$ is our vineyard case study (Section~\ref{sec:example})
\end{scriptsize}
\vspace{-5mm}
\end{table*}